\documentclass[fleqn,usenatbib]{mnras}

\usepackage{newtxtext,newtxmath}

\usepackage[T1]{fontenc}

\DeclareRobustCommand{\VAN}[3]{#2}
\let\VANthebibliography\thebibliography
\def\thebibliography{\DeclareRobustCommand{\VAN}[3]{##3}\VANthebibliography}

\usepackage{graphicx}	
\usepackage{amsmath}	

\newcommand{\Gpc}{\ifmmode  {\rm~Gpc}  \else ${\rm~Gpc}$\fi}
\newcommand{\Gyr}{\ifmmode  {\rm~Gyr}  \else ${\rm~Gyr}$\fi}
\newcommand{\Mpc}{\ifmmode  {\rm~Mpc}  \else ${\rm~Mpc}$\fi}
\newcommand{\kpc}{\ifmmode  {\rm~kpc}  \else ${\rm~kpc}$\fi}
\newcommand{\pkpc}{\ifmmode  {\rm~pkpc}  \else ${\rm~pkpc}$\fi}
\newcommand{\Msun}{\ifmmode {\rm M}_{\odot} \else ${\rm M}_{\odot}$ \fi} 
\newcommand{\Msunpyr}{\ifmmode M_{\odot}{\rm~yr}^{-1} \else $M_{\odot}{\rm~yr}^{-1}$ \fi} 
\newcommand{\kms}{\ifmmode {\rm~km\,s}^{-1} \else ${\rm~km\,s}^{-1}$ \fi} 
\newcommand{\dv}{\ifmmode  {\Delta v}  \else ${\Delta v}$ \fi}
\newcommand{\dr}{\ifmmode  {\Delta r}  \else ${\Delta r}$ \fi}
\newcommand{\drp}{\ifmmode  {\Delta r_{\rm~proj}}  \else ${\Delta r_{\rm~proj}}$ \fi}
\newcommand{\tobs}{\ifmmode  {T_{\rm~obs}}  \else ${T_{\rm~obs}}$ \fi}
\newcommand{\tmerge}{\ifmmode  {T_{\rm~merge}}  \else ${T_{\rm~merge}}$ \fi}
\newcommand{\cmerge}{\ifmmode  {C_{\rm~merge}}  \else ${C_{\rm~merge}}$ \fi}
\newcommand{\rinner}{\ifmmode  {r_{\rm inner}}  \else ${r_{\rm inner}}$ \fi}
\newcommand{\router}{\ifmmode  {r_{\rm outer}}  \else ${r_{\rm outer}}$ \fi}

\newcommand{\dd}[1]{{\mathrm{d}{#1}}}

\title[Merging timescales with {\normalfont \textsc{Emerge}}]{{\normalfont \textsc{Emerge}}: Constraining merging probabilities and timescales of close galaxy pairs}

\author[J. A. O'Leary et al.]{
Joseph A. O'Leary,$^{1}$\thanks{E-mail: joleary@usm.lmu.de}
Benjamin P. Moster,$^{1,2}$
Eva Kr{\"a}mer,$^{3}$
\\
$^{1}$Universit{\"a}ts-Sternwarte, Ludwig-Maximilians-Universit{\"a}t M{\"u}nchen, Scheinerstr. 1, 81679 M{\"u}nchen, Germany\\
$^{2}$Max-Planck Institut f{\"u}r Astrophysik, Karl-Schwarzschild Stra{\ss}e 1, 85748 Garching, Germany\\
$^{3}$Friedrich-Alexander-Universit{\"a}t Erlangen-N{\"u}rnberg, Schlo{\ss}platz 4, 91054 Erlangen, Germany
}

\date{Accepted XXX. Received YYY; in original form ZZZ}

\pubyear{2021}

\begin{document}
\label{firstpage}
\pagerange{\pageref{firstpage}--\pageref{lastpage}}
\maketitle

\begin{abstract}
Theoretical models are vital for exploring the galaxy merger process, which plays a crucial role in the evolution of galaxies. Recent advances in modelling have placed tight constraints on the buildup of stellar material in galaxies across cosmic time. Despite these successes, extracting the merger rates from observable data remains a challenge. Differences in modelling techniques, combined with limited observational data, drive conflicting conclusions on the merging timescales of close pairs. We employ an empirical model for galaxy formation that links galaxy properties to the growth of simulated dark matter halos, along with mock lightcone galaxy catalogues, to probe the dependencies of pair merging probabilities and merging timescales. In this work, we demonstrate that the pair merging probabilities are best described by a logistic function and that mean merging timescales can be well approximated by a linear relation in the projected separation and line of sight velocity difference in observed pairs. Together, our fitting formulae can accurately predict merger rates from galaxy pairs to at least $z\sim4$ under a wide variety of pair selection criteria. Additionally, we show that some commonly used pair selection criteria may not represent a suitable sample of galaxies to reproduce underlying merger rates. Finally, we conclude from our analysis that observation timescales are primarily driven by dynamics and are not strongly impacted by the star formation properties of the component galaxies.
\end{abstract}




\section{Introduction}
Galaxy mergers play a crucial role in the buildup of stellar material under the current hierarchical view of galaxy formation. The galaxy merger process is responsible for not only stellar mass growth, but is also invoked to explain many observed phenomena, such as AGN \citep{Choi2018, Steinborn2018, Gao2020, Marian2020, Sharma2021}, stellar streams, disturbed morphologies \citep{Conselice2003, Lotz2008, Lotz2010, Wen2016, Martin2018, Bluck2019, Mantha2019, Yoon2020}, and quenching \citep{Khalatyan2008, Jesseit2009, Bois2011, Moody2014, Naab2014}. Due to the large timescales involved in galaxy evolution, large volume simulations are necessary to understand vital aspects of galaxy formation such as the galaxy merger rate. In order to validate these theoretical predictions we require theoretical methods to determine these key quantities in an observational context.

When extracting the merger rate of galaxies from observations, the so called observation timescale \tobs is of central importance. This parameter specifies how long a galaxy pair will remain observable under some specified selection criteria. Despite the importance of this value, theoretical models have not yet converged on how this quantity should scale with redshift, stellar mass, mass ratio, projected separation, or redshift proximity \citep{Kitzbichler2008, Lotz2008, Lotz2011, Jiang2014, Snyder2017, Oleary2021}. Efforts to constrain this parameter are hampered by lack of consensus regarding which pair selection criteria should be used, as well as observable limitations which force various groups to adopt different stellar mass, and mass ratio constraints for their analysis \citep{Lin2008, Bundy2009, Man2016, Ventou2017, mantha2018, Duncan2019, Ventou2019}.

Large volume cosmological simulations are a useful tool for probing galaxy growth as they offer a robust statistical sample of galaxies across large dynamic ranges. Different models for galaxy formation offer unique strengths for probing various aspects of the galaxy formation. \textit{Hydrodynamical simulations} are well equipped to explore the impact of baryons on the merging process and built up of stellar material \citep{Dubois2014, Hirschmann2014, Vogelsberger2014b, Schaye2015, Pillepich2018, Hopkins2018}. However, hydrodynamical simulations are computationally expensive, limiting the volumes that can be simulated, and are subject to often uncertain sub-grid models which can obscure the salient aspects of the problem at hand. Alternatively, \textit{Semi-analytic models} (SAMs) offer a more computationally efficient method where galaxies are populated into dark matter halos according to analytic prescriptions \citep{Bower2006, Somerville2008, Benson2012, Henriques2015}. Both these models can struggle to reproduce a large number of observations simultaneously due to the cost of exploring their large sub-grid parameter space.

\textit{Empirical models} of galaxy formation offer a compelling alternative to hydrodynamical sims and SAMs, while providing several distinct advantages \citep{Conroy2009, Moster2010, Behroozi2013d, Moster2013, Moster2018, Behroozi2019, Moster2020}. These models avoid baryonic and sub-grid pitfalls by constructing simplified formulae to relate galaxies to the properties of their host dark matter haloes. Model parameters are then constrained directly by observable data. This approach results in mock galaxy catalogues that meet the relevant observables by design. Modern tree based methods can additionally provide a self-consistent framework for galaxy growth, where scaling relations can be constrained out to high redshift. Moreover, the simplicity and efficiency of these models allows for a thorough exploration of parameter space. A more thorough comparison of our empirical approach to \textit{ab-initio} models can be found in \citet{Moster2018}.

The intent of this work is to further explore the observation timescales introduced in \citet{Oleary2021} to determine more precisely what drives this quantity. To that end we employ the empirical model \textsc{Emerge}, as we can readily translate pair fractions into rates as both can easily be extracted. These can be used to place tighter constraints on \tobs, which can be used to determine rates from observed pairs. Through this work we will provide utilities for observers to translate observed pair fractions into galaxy merger rates. In this effort we will answer three key questions concerning observed merger rates.
\begin{itemize}
    \item What is the probability that two galaxies observed as a close pair will merge by $z=0$ and what is that dependency on radial projected separation\dr, line of sight velocity difference \dv and $z$?
    \item On what timescale will an observed pair merge, and for how long is that pair observable given the pair selection criteria?
    \item What determines pair observation timescales?
\end{itemize}

This paper is organised as follows: In Section~\ref{sec:emerge} we will outline the basic functionality of the empirical model \textsc{emerge}, and describe the underlying $N$-body simulation. In Section~\ref{sec:mocks} we discuss how we construct the mock catalogues that we use in the course of this analysis. Our pair selection criteria and approach to fitting functions are described in Section~\ref{sec:fitting}, with fitting formulas for pair merging probability and merging timescales explain in Section~\ref{sec:C_fit} and Section~\ref{sec:T_fit} respectively. In Section~\ref{sec:rates} we show that the results of Section~\ref{sec:fitting} can be used to reconstruct the merger rates shown in \citet{Oleary2021} when applied to mock catalogues. Finally, in Section~\ref{sec:conclusions} we discuss caveats of our analysis and summarise our key conclusions.

\section{DARK MATTER SIMULATIONS AND EMERGE}
\label{sec:emerge}
Our analysis  relies on producing galaxy merger trees encompassing a large dynamic range, occupying an appropriately large cosmic volume. We employ the empirical model \textsc{emerge} to populate dark matter haloes with galaxies based on individual halo growth histories. The details of this model have been thoroughly discussed in previous works utilising this code \citep{Moster2018, Moster2020, Oleary2021}. In this section we will briefly summarise the aspects of the model most relevant to this work.

\subsection{The simulation}
We utilise a cosmological dark matter only $N$-body simulation in a periodic box with side lengths of $200$ Mpc. This simulation adopts Planck $\Lambda$CDM cosmology \citep{Planck2016} where $\Omega_m = 0.3070$, $\Omega_{\Lambda} = 0.6930$, $\Omega_b = 0.0485$, where $H_0 = 67.77\,\mathrm{km}\,\mathrm{s}^{-1}\mathrm{Mpc}^{-1}$, $n_s=0.9677$, and $\sigma_{8}=0.8149$. The initial conditions for this simulation were generated using \textsc{Music} \citep{music} with a power spectrum obtained from \textsc{CAMB} \citep{camb}. The simulation contains $1024^3$ dark matter particles with particle mass $2.92\times10^8\mathrm{M}_{\odot}$. The simulation was run from $z=63$ to $0$ using the Tree-PM code \textsc{Gadget3} \citep{gadget2}. In total 94 snapshots were created evenly spaced in scale factor $(\Delta a = 0.01)$. Dark matter haloes are identified in each simulation snapshot using the phase space halo finder, \textsc{Rockstar} \citep{rockstar}. Halo merger trees are constructed using \textsc{ConsistentTrees} \citep{ctrees}, providing detailed evolution of physical halo properties across time steps.

\subsection{Emerge in a nutshell}
\textsc{emerge} \citep{Moster2018} takes halo-halo merger trees as an input and populates each halo with a galaxy by linking the galaxy star formation rate (SFR) to the halo growth rate 
\begin{equation}
	\frac{\dd m_*(M, z)}{\dd t} = \frac{\dd m_{\mathrm{bary}}}{\dd t} \epsilon(M,z) = f_{\mathrm{bary}} \frac{\dd M}{\dd t}\epsilon(M,z) \;,
\end{equation}
where $f_b\equiv\Omega_b/\Omega_m$ is the baryon fraction, $\dot{M}$ is the halo growth, $\dot{m}_{\mathrm{bary}}(M,z)$ is the baryonic growth rate which describes how much baryonic material is becoming available, $\epsilon(M,z)$ is the instantaneous conversion efficiency, which determines how efficiently this material can be converted into stars, and $\dot{m}_{*}$ is the SFR.

The conversion efficiency is the core of the model, capturing halo mass and redshift dependent mechanisms that regulate star formation,
\begin{equation}
	\epsilon (M,z) = 2\, \epsilon_N \left[ \left(\frac{M}{M_{1}}\right)^{-\beta}+\left( \frac{M}{M_{1}}\right)^{\gamma}\right]^{-1}  \;,
	\label{eq:efficiency}
\end{equation}
where the normalisation $\epsilon_N$, the characteristic mass $M_1$, and the low and high-mass slopes $\beta$ and $\gamma$ are the free parameters used for the fitting. Furthermore, the model parameters are linearly dependent on the scale factor:
\begin{eqnarray} \label{eq:Mz}
	\log_{10} M_{1}(z) &=& M_0 + M_z\frac{z}{z+1} \;,\\
	\epsilon_N &=&  \epsilon_0 + \epsilon_z\frac{z}{z+1}\;, \\
	\beta(z) &=& \beta_0 + \beta_z\frac{z}{z+1}\;, \\
	\gamma(z) &=& \gamma_0 \;.
\end{eqnarray}
These parameters are allowed to vary freely within their boundary conditions in order to produce a fit in agreement with observation. Observables are chosen such that model parameters can be isolated and independently constrained, thus avoiding degeneracy. For a complete description of the observations used to constrain the model as well as the fitting procedure we refer the reader to \citet{Moster2018, Moster2020} and \citet{Oleary2021}.  The model parameters used in this work are repeated in Table~\ref{tab:best_fit} for completeness. Next we will cover two specific aspects of the model that directly relate to the results shown in this work. 

\begin{table}
	\centering
	\caption{The best fit model parameters used for this work.}
	\label{tab:best_fit}
	\begin{tabular}{lccr} 
		\hline
		Parameter    & Best-fit  & Upper $1\sigma$ & lower $1\sigma$ \\
		\hline
		\hline
		$M_0$        & 11.34829  & +0.03925        & -0.04153        \\
		$M_z$        & 0.654238  & +0.08005        & -0.07242        \\
		$\epsilon_0$ & 0.009010  & +0.00657        & -0.00451        \\
		$\epsilon_z$ & 0.596666  & +0.02880        & -0.02366        \\
		$\beta_0$    & 3.094621  & +0.15251        & -0.14964        \\
		$\beta_z$    & -2.019841 & +0.22206        & -0.20921        \\
		$\gamma_0$   & 1.107304  & +0.05880        & -0.05280        \\
		\hline
		$f_{esc}$    & 0.562183  & +0.02840        & -0.03160        \\
		$f_{s}$      & 0.004015  & +0.00209        & -0.00141        \\
		$\tau_{0}$   & 4.461039  & +0.42511        & -0.40187        \\
		$\tau_{s}$   & 0.346817  & +0.04501        & -0.04265        \\
		\hline
	\end{tabular}
\end{table}

\subsubsection{Galaxy merging}
Merging in this model occurs between galaxies residing at the centre of a resolved $N$-body halo and so-called orphan galaxies. Orphan galaxies are those systems whose host halo has fallen below the resolution limit of the halo finder due to real mass stripping in the simulation. Rather than remove/merge these systems from the simulation when their halo is lost, we continue to track these galaxies within the empirical model using approximate formulae to update their halo mass and position within their host halo system.

The orbits of these orphan galaxies will continue to decay and we merge them with their host system according to some dynamical friction formula. In \citet{Oleary2021} we showed that our merger rates are not strongly driven by our choice of dynamical friction formulation, or our treatment of orphans. This formulation does however play a role in how we update the position of orphan galaxies in emerge, and is thus relevant to the discussion of pair fractions, and their merging timescales.

When a galaxy first becomes an orphan, a dynamical friction clock is set. We use its last known orbital parameters of the orphan's halo to compute the dynamical friction time. We employ the dynamical friction formulation specified by \citet{Boylan-Kolchin2008} to control orphan orbital decay:
\begin{equation}
	t_{\mathrm{df}}=0.0216H(z)^{-1}\frac{(M_{0}/M_{1})^2}{\ln(1+M_{0}/M_{1})}\exp(1.9\eta) \left( \frac{r_{1}}{r_{\mathrm{vir}}}\right)^{2} \;,
	\label{eq:tdf}
\end{equation}
where $H(z)$ is the Hubble parameter, $r_{\mathrm{vir}}$ is the virial radius of the main halo ($M_0$), $r_1$ is the radial position of the subhalo ($M_1$) with with respect to the centre of the main halo, and $\eta$ is a measure for the orbital circularity of the subhalo. This formulation is tuned to high resolution idealised hydrodynamical simulations, and thus includes mass loss due tidal effects and baryonic processes in the merging timescale. When the systems are finally merged stellar mass is added to the descendant system as $m_{\mathrm{desc}} = m_{\mathrm{main}} + m_{\mathrm{orphan}}(1-f_{esc})$ where $m_{\mathrm{desc}}$ is stellar mass of the descendant galaxy, $m_{\mathrm{main}}$ is the stellar mass of the main progenitor galaxy, $m_{\mathrm{orphan}}$ is the orphan galaxy stellar mass, and $f_{esc}$ is the fraction of the orphan stellar mass that will be distributed to the ICM.

\subsubsection{Galaxy clustering}
The focus of this work involves galaxies in close pairs. The galaxy pair fraction is related to, but distinct from, the projected galaxy correlation function ($w_{p}$), which is one of the observations used to constrain \textsc{emerge}. Galaxy clustering in this model is largely driven by our implementation for satellite stripping. The stripping model in \textsc{emerge} is a simple halo mass threshold
\begin{equation}
	M < f_s \, M_{\mathrm{peak}}\; ,
\end{equation}
where $M$ is the current halo mass, $M_{\mathrm{peak}}$ is the peak halo mass, and $f_s$ is the stripping fraction. In the case of orphans we estimate their current halo mass by assuming mass is stripped at a constant rate defined between the time of peak mass and the time the halo was lost in the simulation \citep[see section 2.5 of][]{Oleary2021}. As a subhalo orbits within its host halo it will gradually lose mass due to tidal stripping. When the halo has fallen below some fraction of $M_{\mathrm{peak}}$ the galaxy residing in that halo will be stripped and its stellar material distributed to the ICM. The consequence of this formula is that lower $f_s$ will drive stronger clustering at small scales as more satellites can survive to reach those small separations. Conversely, larger $f_s$ will reduce clustering by stripping galaxies sooner.

Finally, in order to compute clustering we need to know the position of each galaxy in the simulation volume, including orphans. For galaxies in resolved haloes we use the position of the $N$-body halo directly to compute clustering. In the case of orphans, positions are approximated by placing the orphan galaxy randomly on a sphere of radius
\begin{equation}
    \label{eqn:radius}
    r=r_0 \sqrt{1-\Delta t /t_{\mathrm{df}}}
\end{equation}
centred on the main halo. Here $r_0$ is the radial position of the orphan when its halo was last resolved and $\Delta t$ is the time elapsed since the subhalo was last resolved \citep{Binney87}.

\section{From simulation to observation}
\label{sec:mocks}
Simulations are a tool needed to interpret observed data. The galaxy-galaxy merger rate is a particularly difficult quantity to derive from the relatively static view of the the universe we see in galaxy surveys. As we cannot view the complete merging of two galaxies in real time a proxy is required as stand in. Obvious physical tracers of a recent merger such as disturbed morphologies present one option for deducing the galaxy merger rate. Methods invoking quantitative morphology such as $G-M_{20}$ or asymmetry are not equally sensitive to all merger mass ratios. Furthermore, these morphological methods are sensitive to total mass, gas properties, orbital parameters, merger stage, and viewing angles \citep[]{Abraham2003, Conselice2003, Lotz2008, Lotz2011, Scarlata2007}.  These additional difficulties present a greater barrier to identifying mergers and determining a cosmological merger rate \citep[]{Kampczyk2007, Scarlata2007, Lopez-Sanjuan2009, Shi2009, Kartaltepe2010, Abruzzo2018, Nevin2019}.

One common observational method for deriving the galaxy merger rate is through the analysis of galaxies in close pairs. The foundation of this approach is simple, as galaxies found in close proximity are expected to merge within some finite predictable time scale. Three key questions must be answered to utilise this method: How should we select galaxy pairs, do those two pairs merge, and on what timescale does an observed pair merge?

\subsection{Light-cone construction}
\label{sec:lightcone}
Throughout this work we will make use of mock light-cone galaxy catalogues. Working with light-cones offers a few advantages to working with snapshot catalogues directly. The first reason is that these catalogues provide a more natural comparison between simulation and observation. In a snapshot catalogue all galaxies exist at the same cosmological stage of evolution. In contrast, real observations have to contend with galaxies samples that span often large redshift ranges encompassing galaxies at various stages of evolution. Our catalogues are constructed in the same way, such that a galaxy is placed at its cosmologically relevant stage of evolution according to the comoving distance from the observer. Another advantage is that these catalogues inherently adopt the same constraints seen in observation due to limited viewing angles. While this does limit the sample size we can use for analysis, it provides a more appropriate environment to test how the models we develop and the conclusions drawn are impacted by these real limitations.

Our cone geometry is set using the method described by \citet{Kitzbichler2008}. In this method light-cone geometry is defined by two integers $m$ and $n$. The line of sight vector $u_3$, is defined by a line drawn from the origin though the point $(L_{\mathrm{box}}/m,L_{\mathrm{box}}/n,L_{\mathrm{box}})$, where $L_{\mathrm{box}}$ is the comoving side length of our simulation volume ($200$\Mpc). The second vector $u_1$ is defined to be orthogonal to $u_3$ and the coordinate axis corresponding to the smaller value of $m$ and $n$. The final vector $u_2$ is defined to be orthogonal to $u_3$ and $u_1$, where all three taken together form a right handed coordinate system. The observation area of he light-cone is then covered by $(m^2n)^{-1}\times(mn^2)^{-1}\mathrm{rad}^2$, which is centred along $u_3$ with edges aligned along $u_1$ and $u_2$.

When galaxies are placed into light cones we draw from snapshots according to
\begin{equation}
\label{eqn:spacing}
    \frac{D_{i} + D_{i-1}}{2} \leq D_{\mathrm{gal}} < \frac{D_{i} + D_{i+1}}{2}
\end{equation}
where $D_i$ is the cosmological distance to simulation snapshot with index $i$, and $D_{\mathrm{gal}}$ is the comoving distance to the galaxy within the light-cone. The `cosmological' redshift of each galaxy is then set by that comoving distance from the observer. Additionally, we apply redshiftspace distortions to each catalogue galaxy to obtain its observed redshift $z_{\mathrm{obs}}$. 

We construct a series of light-cones intended to reproduce the observation area of the five CANDLES \citep{Koekemoer2011} survey fields. These fields serve as the test bed for applying our fitting formula. In addition to these mock catalogues we construct an additional `full width' light-cone catalogue. This data-set has galaxies placed according to eq.~\ref{eqn:spacing} just as in the standard light-cone catalogues. However the line of sight vector $u_3$ is aligned with the coordinate $z$-axis, and no view restricting cone geometry is applied. This results in a catalogue with galaxies occupying a rectangular volume and creates a `smoothly' evolving galaxy catalogue without the restrictions due to limited viewing angle, which is helpful when fitting data at low redshift. 

\subsection{Identifying close pairs}
\begin{figure}
	\includegraphics[width=\columnwidth]{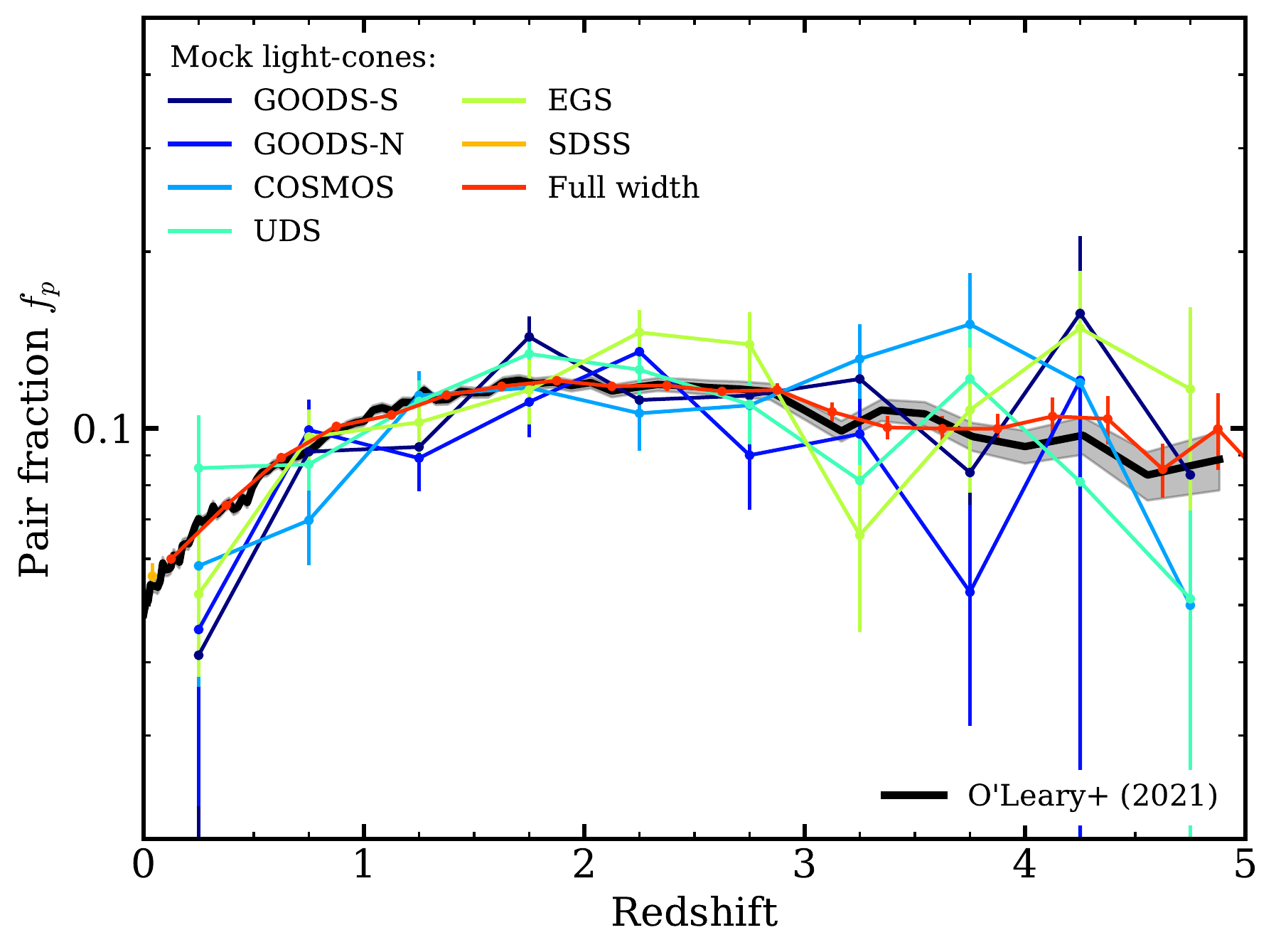}
	\caption{The pair fraction evolution in our simulated galaxy catalogues for pairs with $\log_{10}(m_{1}/M_{\odot}) \geq 10.3$,  $5\leq\dr<50\kpc$ projected separation, $\dv \leq 500\kms$ and $1\leq\mu<4$. The solid black line indicates the pair fraction computed at each simulation snapshot as in \citet{Oleary2021}. Coloured lines show the pair fraction evolution with our mock light-cones. Poisson error in the number count of pairs is reflected in the error bars for the mock light-cones, and the grey shaded region for the results of \citet{Oleary2021}.} Here we do not place any redshift restraints on the light cone catalogues that would more closely resemble the observables limits of the noted surveys.
	\label{fig:pair_fraction}
\end{figure}

\begin{figure*}
	\includegraphics[width=\textwidth,height=\textheight,keepaspectratio]{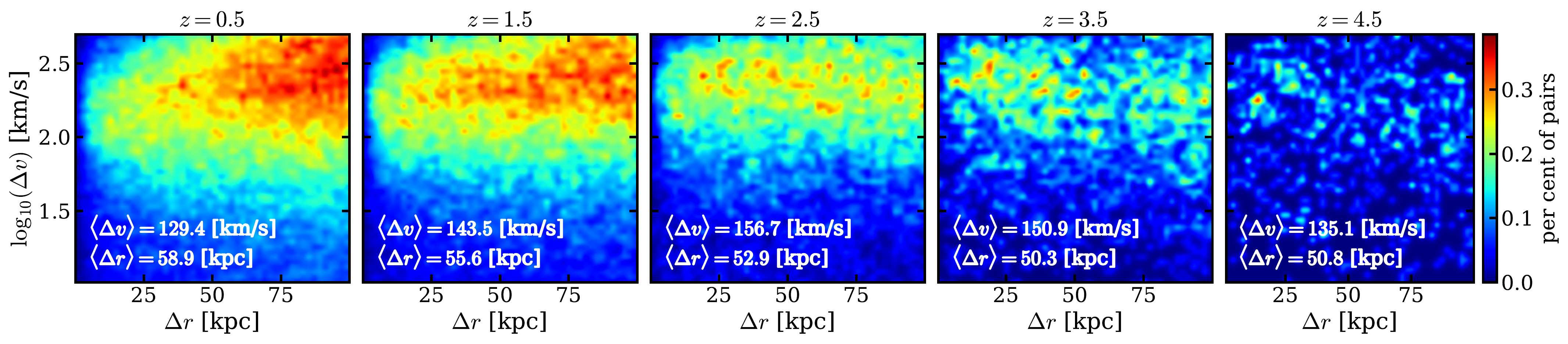}
	\caption{The distribution of pairs for $\log_{10}(m_1) \geq 10.3$ and $1 \leq \mu < 4$. Each panel includes pairs where the observed redshift of the primary galaxy falls within $\pm0.5$ of the noted central redshift. The percentages shown correspond to pairs within that redshift bin, not percentages to the entire catalogue of pairs.}
	\label{fig:pair_dist}
\end{figure*}

In this work pairs are identified and defined in terms of the following:
\begin{description}
	\item $m_1$: The stellar mass of the \textit{main} (more massive) galaxy in each pair.
	\item $\mu$: The stellar mass ratio taken with respect to the two galaxies forming the pair, $\mu \equiv m_1/m_2$. Here $m_2$ is the \textit{minor} (less massive) galaxy in the pair. The mass ratio is defined such that $\mu \geq 1$.
	\item $z$: Pair redshift measured at the observed redshift of the main galaxy.
	\item \dr: The \textit{projected} radial separation between the two galaxies in \pkpc.
	\item \dv: The line-of-sight velocity difference between the two galaxies, as measured by the difference in their respective $z_{\mathrm{obs}}$
\end{description}
In this work we provide fits for a variety of mass and mass ratio combinations. For simplicity of conveying key concepts and conclusions we use a single reference case when displaying results. Which refers to pairs with $\log_{10}(m_1) \geq 10.3,\;\dv \leq 500 \kms$ and $1\leq\mu<4$. This selection criteria is used for ease of comparison with \citet{Oleary2021}. In Figure~\ref{fig:pair_fraction} we compare the pair fractions determined from the data sets described in Section~\ref{sec:lightcone} with the pair fractions of \citet{Oleary2021}, which computed the pairs on a snapshot by snapshot basis. With this we are able to see that our underlying results agree, and can get a good idea over the amount of uncertainty in the pair fraction under more realistic observable constraints. A more complete overview of how pairs are distributed in our simulation volume is shown in Figure \ref{fig:pair_dist}.

\section{Merging probabilities and timescales}
\label{sec:fitting}
Once we have a handle on the pair fraction, we can determine the galaxy merger rates: for a given mass range, observable aperture and redshift proximity criteria we compute the pair fraction and divide that value by the average time that pairs under that selection criteria will be observable. The galaxy rate can then be expressed as:
\begin{equation}
    \label{eqn:rate1}
    \mathfrak{R} = C_{\mathrm{merge}} \frac{f_{p}}{\langle T_{\mathrm{obs}} \rangle}\; ,
\end{equation}
where $\langle \tobs \rangle$ is the average observation timescale. The term $C_\mathrm{merge}$ is an optional correction factor to account for pairs that do not merge before $z=0$ or at all. In general these two quantities are dependent on the pair selection criteria imposed by the observer.

The goal of this work is to characterise these values, and provide meaningful formulations that reduce the need to establish fitting functions for each specific observation. In the theoretical framework there are a few ways we can do this. One approach is to use idealised galaxy merger simulations, making mock pair observations and tracking these merging pairs to final coalescence \citep{Lotz2011}. Alternatively, one could employ some self-consistent model for galaxy formation in large volumes, and track pairs identified in mock observations \citep{Snyder2017}. Previous efforts have focused on setting a fixed pair selection criteria, and finding some fitting function for the observation timescale under that criteria. The issue with that approach is that those fitting functions cannot be easily applied to pairs identified under some other criteria. 

In the next two sections we provide fitting functions for both \cmerge and \tmerge. We will describe the process we use to determine these formulae that can describe these values for a range of \dr, \dv, and redshift commonly used in the literature. In both cases parameter space is explored and best fit values determined using the affine invariant ensemble sampler described in \citet{Goodman2010} as implemented in \texttt{Emcee} \citep{Foreman-Mackey2013}. For each mass range explored we fit to three mass ratio intervals $1:4$ (major), $4:10$ (minor), and $1:10$ (all). Throughout, we only included pairs with $\dv\leq 500 \kms$ and $\dr\leq 100\kpc$. 

As an aside, the nature of this problem makes it attractive to deal with from a machine learning perspective. If our central interest is determining the observation timescale of any pair regardless of the mechanisms driving those timescales it would appear on its surface to be an ideal problem for machine learning algorithms. We tested this approach using a random forest regressor to predict merging probabilities and timescales. In practice we found there was not enough information in the \textit{observable} pair features to meaningfully predict the desired values on an individual basis, due to information loss in projected quantities. Other works have had greater success on this front using more advanced networks \citep{Pfister2020}.

\subsection{Merging Probability}
\label{sec:C_fit}
\begin{figure*}
	\includegraphics[width=\textwidth,height=\textheight,keepaspectratio]{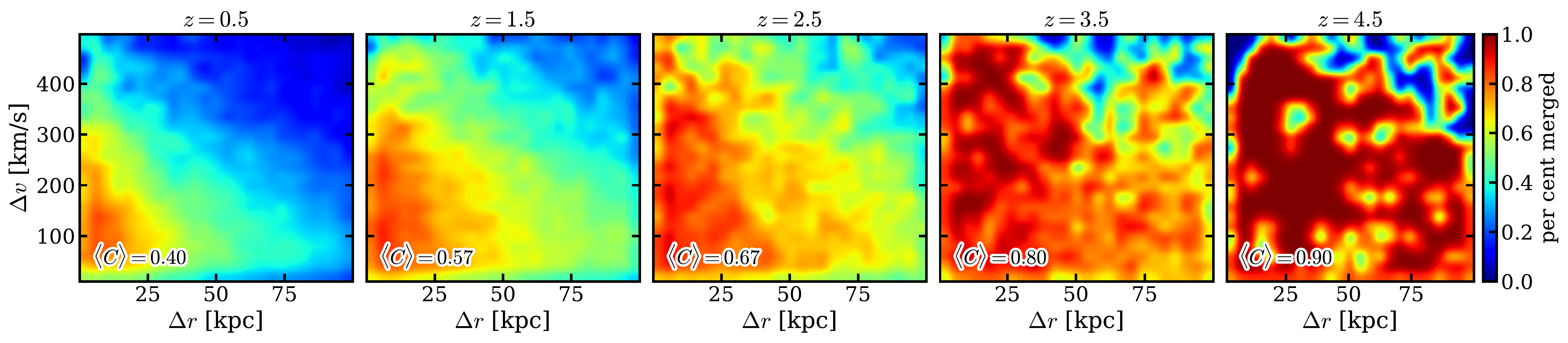}
	\caption{The per cent of pairs that merge by $z=0$ for pairs with $\log_{10}(m_1) \geq 10.3$ with $1 \leq \mu < 4$. Each panel includes pairs where the observed redshift of the primary galaxy falls within $\pm0.5$ of the noted central redshift.}
	\label{fig:pair_prob}
\end{figure*}
First we need to address the merging probability. Traditionally chance pairs, and pairs that did not have enough time to merge with in the average timescale were captured through a correction factor $C_\mathrm{merge}$. This correction factor typically takes some value between $0.4$ and $1.0$. For instance \citet{Lotz2011} adopt a constant $0.6$ for all scenarios they tested. Other works have chosen to marignalize over this parameter by including it directly into the $T_{\mathrm{obs}}$ formulation. This the approach taken in \citet{Snyder2017} as well as in \citet{Oleary2021}.

Several recent works have adopted a probabilistic approach \citep{Duncan2019, Ventou2019}. In these works each pair is assigned a weight, based on some observed properties, that the pair is physically associated and will merge on in the expected timescale. \citet{Ventou2019} derived their weighting function through pairs selected in the Illustris simulation, in their results they determined the merger probability could be well described by an exponential function in both \dr and \dv with little redshift dependence. Looking at Figure~\ref{fig:pair_prob} we can immediately see a redshift dependent formulation is required. Just probing this coarse redshift bins shown in Figure~\ref{fig:pair_prob} we find the probability of merging ranges between $\sim0.4$ and $\sim1.0$. For this selection criteria our merger probability appears to saturate near $z\approx3.5$. However it should not be surprising that there is a redshift dependence to the merger probability. This value is defined as the probability that two galaxies will merge by $z=0$ which directly implies that a pair at $z=0$ would have no chance to merge. Similarly, if two pairs at high z are in fact physically associated, there is simply more time available where they \textit{could} merge before present day. Redshift dependencies not withstanding, we found that the fitting formula of \citet{Ventou2019} can only reasonably fit our data for $z\lesssim 1$.

After testing several candidate fitting functions we found that the merger probability can be well described by a logistic function in velocity, with redshift and radial dependencies, eq.~\ref{eqn:C}.
\begin{align}
  W(\dr,\dv,z) & = \frac{\exp(b \dr)}{1 + \exp[c_{0}(\dv-a)]} \label{eqn:C} \\
    a & = a_0(1+z)^{a_z} + a_r\dr \nonumber \\
    b & = b_0 + (1+z){b_z} \nonumber
\end{align}
 In this formulation the maximum of the curve is set by the term $\exp(b\dr)$ where the exponential slope $b$ has a linear dependence on redshift. The logistic midpoint $a$ was found to have dependencies in both redshift and \dr. For this parameter a linear relation to \dr and a power-law relation to redshift produced the best results. We did not find that the logistic growth rate $c_{0}$, varies strongly with either \dr or redshift. Eq.~\ref{eqn:C} reproduces the data best for smaller \dr and \dv. 
 
 Table~\ref{tab:W_fit} Shows the best fit parameters from eq.~\ref{eqn:C} for a range of primary stellar mass thresholds, and mass ratios. For each mass range we fit to major ($1\leq \mu <4$), minor ($4\leq \mu <10$) and major+minor ($1\leq \mu <10$) pair mass ratios.
 
\begin{table*}
\caption{Best fit parameters for merging probability eq.~\ref{eqn:C}}
\label{tab:W_fit}
    \begin{tabular}{lccccccr}
        \hline
        $\log_{10}(m_{1}/\Msun)$    & Mass ratio $\mu$  & $a_{0}\, [\kms]$                & $a_{z}$                & $a_{r}\, [\kms\kpc^{-1}]$                 & $b_{0}\, [\kpc^{-1}]$                       & $b_{z}\, [\kpc^{-1}]$                     & $c_{0}\, [\mathrm{s}\, \mathrm{km}^{-1}]$                       \\
        \hline
        \hline
                                    & $1 \leq \mu<4$    & $41.1^{+16.8}_{-27.2}$ & $1.20^{+0.35}_{-0.35}$ & $-1.16^{+1.31}_{-1.08}$ & $-0.0277^{+0.0078}_{-0.0098}$ & $0.0049^{+0.0020}_{-0.0016}$ & $0.0083^{+0.0013}_{-0.0016}$ \\
        $9.0 - 10.0$                & $4 \leq \mu<10$   & $72.5^{+29.3}_{-36.1}$ & $0.95^{+0.30}_{-0.35}$ & $-1.43^{+1.13}_{-0.98}$ & $-0.0248^{+0.0096}_{-0.0124}$ & $0.0033^{+0.0026}_{-0.0021}$ & $0.0108^{+0.0019}_{-0.0025}$ \\
                                    & $1 \leq \mu<10$   & $57.6^{+23.6}_{-31.6}$ & $1.06^{+0.31}_{-0.35}$ & $-1.28^{+1.15}_{-0.97}$ & $-0.0265^{+0.0086}_{-0.0108}$ & $0.0040^{+0.0022}_{-0.0018}$ & $0.0098^{+0.0016}_{-0.0020}$ \\
        \hline
                                    & $1 \leq \mu<4$    & $234^{+49}_{-48}$ & $0.629^{+0.167}_{-0.210}$ & $-1.84^{+1.25}_{-1.00}$ & $-0.0141^{+0.0056}_{-0.0065}$ & $0.0021^{+0.0014}_{-0.0012}$ & $0.0084^{+0.0021}_{-0.0029}$ \\
        $10.0 - 11.0$               & $4 \leq \mu<10$   & $197^{+47}_{-48}$ & $0.665^{+0.176}_{-0.221}$ & $-2.00^{+1.18}_{-0.95}$ & $-0.0145^{+0.0068}_{-0.0079}$ & $0.0019^{+0.0017}_{-0.0015}$ & $0.0093^{+0.0022}_{-0.0029}$ \\
                                    & $1 \leq \mu<10$   & $215^{+44}_{-44}$ & $0.644^{+0.153}_{-0.192}$ & $-1.89^{+1.13}_{-0.92}$ & $-0.0147^{+0.0057}_{-0.0067}$ & $0.0021^{+0.0013}_{-0.0012}$ & $0.0089^{+0.0020}_{-0.0027}$ \\
        \hline
                                    & $1 \leq \mu<4$    & $400^{+120}_{-217}$ & $1.17^{+0.43}_{-0.41}$ & $-1.58^{+2.20}_{-2.08}$ & $-0.0141^{+0.0041}_{-0.0042}$ & $0.0027^{+0.0011}_{-0.0011}$ & $0.0030^{+0.0014}_{-0.0026}$ \\
        $11.0 - 12.0$               & $4 \leq \mu<10$   & $152^{+43}_{-60}$   & $1.33^{+0.36}_{-0.31}$ & $-0.54^{+2.20}_{-1.48}$ & $-0.0264^{+0.0062}_{-0.0067}$ & $0.0045^{+0.0016}_{-0.0015}$ & $0.0050^{+0.0017}_{-0.0026}$ \\
                                    & $1 \leq \mu<10$   & $339^{+110}_{-211}$ & $1.12^{+0.40}_{-0.41}$ & $-1.33^{+2.30}_{-2.02}$ & $-0.0170^{+0.0045}_{-0.0046}$ & $0.0031^{+0.0011}_{-0.0011}$ & $0.0026^{+0.0012}_{-0.0020}$ \\
        \hline
                                    & $1 \leq \mu<4$    & $238^{+66}_{-71}$ & $0.467^{+0.200}_{-0.248}$ & $-2.38^{+1.47}_{-1.35}$ & $-0.0165^{+0.0067}_{-0.0073}$ & $0.0029^{+0.0015}_{-0.0013}$ & $0.0057^{+0.0012}_{-0.0017}$ \\
        $\geq 9.5$                  & $4 \leq \mu<10$   & $163^{+45}_{-47}$ & $0.716^{+0.196}_{-0.245}$ & $-2.17^{+1.20}_{-1.04}$ & $-0.0165^{+0.0081}_{-0.0092}$ & $0.0021^{+0.0019}_{-0.0017}$ & $0.0086^{+0.0017}_{-0.0024}$ \\
                                    & $1 \leq \mu<10$   & $200^{+55}_{-59}$ & $0.584^{+0.200}_{-0.246}$ & $-2.18^{+1.43}_{-1.23}$ & $-0.0170^{+0.0074}_{-0.0080}$ & $0.0026^{+0.0016}_{-0.0015}$ & $0.0068^{+0.0014}_{-0.0020}$ \\
        \hline
                                    & $1 \leq \mu<4$    & $262^{+64}_{-68}$ & $0.494^{+0.186}_{-0.230}$ & $-2.35^{+1.46}_{-1.30}$ & $-0.0151^{+0.0063}_{-0.0069}$ & $0.0025^{+0.0014}_{-0.0013}$ & $0.0061^{+0.0013}_{-0.0020}$ \\
        $\geq 9.7$                  & $4 \leq \mu<10$   & $176^{+45}_{-47}$ & $0.718^{+0.184}_{-0.228}$ & $-2.14^{+1.25}_{-1.03}$ & $-0.0159^{+0.0074}_{-0.0085}$ & $0.0021^{+0.0017}_{-0.0015}$ & $0.0085^{+0.0018}_{-0.0024}$ \\
                                    & $1 \leq \mu<10$   & $218^{+55}_{-58}$ & $0.604^{+0.186}_{-0.232}$ & $-2.27^{+1.40}_{-1.21}$ & $-0.0153^{+0.0067}_{-0.0074}$ & $0.0023^{+0.0015}_{-0.0014}$ & $0.0068^{+0.0015}_{-0.0021}$ \\
        \hline
                                    & $1 \leq \mu<4$    & $282^{+65}_{-69}$ & $0.592^{+0.195}_{-0.237}$ & $-2.39^{+1.51}_{-1.41}$ & $-0.0131^{+0.0056}_{-0.0061}$ & $0.0021^{+0.0013}_{-0.0012}$ & $0.0060^{+0.0014}_{-0.0023}$ \\
        $\geq 10.0$                 & $4 \leq \mu<10$   & $188^{+45}_{-47}$ & $0.752^{+0.184}_{-0.229}$ & $-2.08^{+1.32}_{-1.12}$ & $-0.0152^{+0.0068}_{-0.0077}$ & $0.0021^{+0.0016}_{-0.0014}$ & $0.0082^{+0.0018}_{-0.0025}$ \\
                                    & $1 \leq \mu<10$   & $249^{+55}_{-57}$ & $0.617^{+0.175}_{-0.212}$ & $-2.41^{+1.38}_{-1.23}$ & $-0.0140^{+0.0058}_{-0.0065}$ & $0.0021^{+0.0013}_{-0.0012}$ & $0.0067^{+0.0015}_{-0.0023}$ \\
        \hline
                                    & $1 \leq \mu<4$    & $289^{+65}_{-71}$ & $0.71^{+0.22}_{-0.27}$ & $-2.10^{+1.65}_{-1.57}$ & $-0.0126^{+0.0050}_{-0.0054}$ & $0.0020^{+0.0012}_{-0.0011}$ & $0.0061^{+0.0017}_{-0.0027}$ \\
        $\geq 10.3$                 & $4 \leq \mu<10$   & $190^{+50}_{-47}$ & $-2.17^{+1.29}_{-1.44}$   & $0.84^{+0.24}_{-0.20}$  & $-0.0144^{+0.0072}_{-0.0064}$ & $0.0020^{+0.0013}_{-0.0015}$ & $0.0074^{+0.0025}_{-0.0017}$ \\
                                    & $1 \leq \mu<10$   & $260^{+58}_{-62}$ & $0.70^{+0.19}_{-0.23}$ & $-2.52^{+1.43}_{-1.40}$ & $-0.0126^{+0.0053}_{-0.0057}$ & $0.0019^{+0.0012}_{-0.0011}$ & $0.0061^{+0.0014}_{-0.0023}$  \\
        \hline
                                    & $1 \leq \mu<4$    & $295^{+67}_{-76}$ & $0.813^{+0.241}_{-0.315}$ & $-1.97^{+1.75}_{-1.67}$ & $-0.0124^{+0.0044}_{-0.0049}$ & $0.0021^{+0.0011}_{-0.0010}$ & $0.0057^{+0.0017}_{-0.0028}$ \\
        $\geq 10.5$                 & $4 \leq \mu<10$   & $175^{+49}_{-54}$ & $0.966^{+0.253}_{-0.309}$ & $-1.52^{+1.70}_{-1.49}$ & $-0.0164^{+0.0065}_{-0.0074}$ & $0.0024^{+0.0016}_{-0.0014}$ & $0.0071^{+0.0018}_{-0.0027}$  \\
                                    & $1 \leq \mu<10$   & $255^{+62}_{-67}$ & $0.816^{+0.222}_{-0.279}$ & $-2.04^{+1.74}_{-1.65}$ & $-0.0133^{+0.0051}_{-0.0055}$ & $0.0020^{+0.0012}_{-0.0011}$ & $0.0057^{+0.0015}_{-0.0024}$  \\
        \hline
                                    & $1 \leq \mu<4$    & $321^{+80}_{-111}$ & $1.04^{+0.34}_{-0.42}$ & $-1.82^{+1.98}_{-1.97}$ & $-0.0128^{+0.0040}_{-0.0045}$ & $0.0023^{+0.0011}_{-0.0010}$ & $0.0044^{+0.0016}_{-0.0027}$ \\
        $\geq 10.8$                 & $4 \leq \mu<10$   & $152^{+44}_{-56}$  & $1.21^{+0.33}_{-0.35}$ & $-0.65^{+2.01}_{-1.46}$ & $-0.0209^{+0.0061}_{-0.0068}$ & $0.0033^{+0.0015}_{-0.0014}$ & $0.0061^{+0.0018}_{-0.0027}$ \\
                                    & $1 \leq \mu<10$   & $267^{+72}_{-93}$  & $1.03^{+0.32}_{-0.39}$ & $-1.53^{+2.08}_{-1.90}$ & $-0.0149^{+0.0044}_{-0.0049}$ & $0.0024^{+0.0011}_{-0.0010}$ & $0.0042^{+0.0015}_{-0.0024}$ \\
        \hline
                                    & $1 \leq \mu<4$    & $400^{+120}_{-220}$ & $1.16^{+0.43}_{-0.41}$ & $-1.51^{+2.22}_{-2.05}$ & $-0.0141^{+0.0040}_{-0.0042}$ & $0.0028^{+0.0011}_{-0.0011}$ & $0.0030^{+0.0014}_{-0.0027}$ \\
        $\geq 11.0$                 & $4 \leq \mu<10$   & $153^{+43}_{-62}$   & $1.33^{+0.36}_{-0.32}$ & $-0.54^{+2.17}_{-1.49}$ & $-0.0264^{+0.0062}_{-0.0068}$ & $0.0045^{+0.0016}_{-0.0015}$ & $0.0050^{+0.0017}_{-0.0027}$ \\
                                    & $1 \leq \mu<10$   & $339^{+112}_{-209}$ & $1.12^{+0.40}_{-0.42}$ & $-1.29^{+2.34}_{-1.98}$ & $-0.0169^{+0.0044}_{-0.0045}$ & $0.0031^{+0.0011}_{-0.0011}$ & $0.0026^{+0.0012}_{-0.0020}$\\
        \hline
    \end{tabular}
\end{table*}
 
\subsection{Merging Timescales}
\label{sec:T_fit}

\begin{table*}
\caption{Best fit parameters for merging timescale eq.~\ref{eqn:T}. }
\label{tab:T_fit}
    \begin{tabular}{lccccccr}
        \hline
        $\log_{10}(m_{1}/\Msun)$    & Mass ratio $\mu$  & $a_{0}\, [\Gyr]$                   & $a_{z}\, [\Gyr]$                   & $b_{0}\, [\Gyr\kpc^{-1}]$                    & $b_{z}\, [\Gyr\kpc^{-1}]$                      & $c_{0}\, [\Gyr\,\mathrm{s}\, \mathrm{km}^{-1}]$                    & $c_{z}\, [\Gyr\,\mathrm{s}\, \mathrm{km}^{-1}]$                        \\
        \hline
        \hline
                                    & $1 \leq \mu<4$    & $0.533^{+0.427}_{-0.420}$ & $0.0312^{+0.0423}_{-0.0420}$ & $0.0426^{+0.0252}_{-0.0252}$ & $-0.0050^{+0.0064}_{-0.0064}$ & $0.0118^{+0.0091}_{-0.0089}$ & $-0.0026^{+0.0021}_{-0.0022}$ \\
        $9.0 - 10.0$                & $4 \leq \mu<10$   & $0.498^{+0.401}_{-0.398}$ & $-0.0597^{+0.0808}_{-0.0797}$ & $0.0343^{+0.0089}_{-0.0090}$ & $-0.0011^{+0.0015}_{-0.0015}$ & $0.0077^{+0.0059}_{-0.0058}$ & $-0.0015^{+0.0014}_{-0.0014}$ \\
                                    & $1 \leq \mu<10$   & $0.483^{+0.372}_{-0.375}$ & $-0.0289^{+0.0385}_{-0.0390}$ & $0.0429^{+0.0202}_{-0.0204}$ & $-0.0040^{+0.0051}_{-0.0051}$ & $0.0081^{+0.0066}_{-0.0065}$ & $-0.0016^{+0.0016}_{-0.0016}$ \\
        \hline
                                    & $1 \leq \mu<4$    & $-0.284^{+0.330}_{-0.330}$ & $0.224^{+0.105}_{-0.105}$ & $0.0315^{+0.0079}_{-0.0079}$ & $-0.0036^{+0.0023}_{-0.0023}$ & $0.0068^{+0.0015}_{-0.0015}$ & $-0.0019^{+0.0004}_{-0.0004}$ \\
        $10.0 - 11.0$               & $4 \leq \mu<10$   & $0.083^{+0.105}_{-0.106}$ & $0.024^{+0.031}_{-0.031}$ & $0.0334^{+0.0093}_{-0.0093}$ & $-0.0028^{+0.0029}_{-0.0029}$ & $0.0069^{+0.0017}_{-0.0017}$ & $-0.0015^{+0.0005}_{-0.0005}$ \\
                                    & $1 \leq \mu<10$   & $-0.164^{+0.211}_{-0.214}$ & $0.146^{+0.078}_{-0.078}$ & $0.0334^{+0.0085}_{-0.0085}$ & $-0.0036^{+0.0026}_{-0.0026}$ & $0.0069^{+0.0016}_{-0.0016}$ & $-0.0017^{+0.0005}_{-0.0004}$ \\
        \hline
                                    & $1 \leq \mu<4$    & $-0.265^{+0.277}_{-0.277}$ & $0.179^{+0.084}_{-0.083}$ & $0.0311^{+0.0053}_{-0.0052}$ & $-0.0058^{+0.0016}_{-0.0016}$ & $0.0038^{+0.0009}_{-0.0009}$ & $-0.0011^{+0.0003}_{-0.0003}$\\
        $11.0 - 12.0$               & $4 \leq \mu<10$   & $0.450^{+0.178}_{-0.176}$  & $-0.041^{+0.050}_{-0.052}$ & $0.0452^{+0.0084}_{-0.0078}$ & $-0.0093^{+0.0025}_{-0.0026}$ & $-0.0005^{+0.0005}_{-0.0005}$ & $0.0003^{+0.0001}_{-0.0002}$ \\
                                    & $1 \leq \mu<10$   & $0.133^{+0.094}_{-0.094}$  & $0.0003^{+0.0005}_{-0.0005}$ & $0.0272^{+0.0051}_{-0.0054}$ & $-0.0032^{+0.0014}_{-0.0015}$ & $0.0030^{+0.0008}_{-0.0008}$ & $-0.0007^{+0.0002}_{-0.0002}$  \\
        \hline
                                    & $1 \leq \mu<4$    & $-0.137^{+0.181}_{-0.181}$ & $0.192^{+0.081}_{-0.082}$ & $0.0315^{+0.0091}_{-0.0092}$ & $-0.0028^{+0.0028}_{-0.0028}$ & $0.0053^{+0.0014}_{-0.0014}$ & $-0.0014^{+0.0004}_{-0.0004}$ \\
        $\geq 9.5$                  & $4 \leq \mu<10$   & $0.135^{+0.161}_{-0.161}$ & $0.027^{+0.035}_{-0.035}$ & $0.0323^{+0.0078}_{-0.0078}$ & $-0.0018^{+0.0022}_{-0.0022}$ & $0.0064^{+0.0021}_{-0.0021}$ & $-0.0014^{+0.0006}_{-0.0006}$ \\
                                    & $1 \leq \mu<10$   & $0.012^{+0.016}_{-0.016}$ & $0.101^{+0.065}_{-0.066}$ & $0.0327^{+0.0092}_{-0.0093}$ & $-0.0025^{+0.0028}_{-0.0028}$ & $0.0055^{+0.0016}_{-0.0016}$ & $-0.0013^{+0.0005}_{-0.0005}$ \\
        \hline
                                    & $1 \leq \mu<4$    & $-0.058^{+0.077}_{-0.077}$ & $0.149^{+0.063}_{-0.063}$ & $0.0276^{+0.0064}_{-0.0063}$ & $-0.0017^{+0.0019}_{-0.0019}$ & $0.0051^{+0.0011}_{-0.0012}$ & $-0.0014^{+0.0004}_{-0.0003}$ \\
        $\geq 9.7$                  & $4 \leq \mu<10$   & $0.094^{+0.121}_{-0.123}$ & $0.034^{+0.042}_{-0.043}$ & $0.0322^{+0.0078}_{-0.0078}$ & $-0.0019^{+0.0023}_{-0.0023}$ & $0.0067^{+0.0018}_{-0.0018}$ & $-0.0015^{+0.0005}_{-0.0005}$  \\
                                    & $1 \leq \mu<10$   & $0.015^{+0.020}_{-0.020}$ & $0.098^{+0.060}_{-0.060}$ & $0.0313^{+0.0083}_{-0.0085}$ & $-0.0024^{+0.0025}_{-0.0025}$ & $0.0053^{+0.0014}_{-0.0014}$ & $-0.0013^{+0.0004}_{-0.0004}$  \\
        \hline
                                    & $1 \leq \mu<4$    & $-0.187^{+0.233}_{-0.238}$ & $0.187^{+0.082}_{-0.082}$ & $0.0291^{+0.0071}_{-0.0070}$ & $-0.0030^{+0.0021}_{-0.0021}$ & $0.0055^{+0.0012}_{-0.0012}$ & $-0.0015^{+0.0004}_{-0.0004}$ \\
        $\geq 10.0$                 & $4 \leq \mu<10$   & $0.102^{+0.124}_{-0.124}$  & $0.025^{+0.031}_{-0.031}$ & $0.0336^{+0.0091}_{-0.0090}$ & $-0.0029^{+0.0027}_{-0.0027}$ & $0.0061^{+0.0016}_{-0.0016}$ & $-0.0013^{+0.0004}_{-0.0004}$  \\
                                    & $1 \leq \mu<10$   & $-0.044^{+0.059}_{-0.059}$ & $0.109^{+0.053}_{-0.054}$ & $0.0309^{+0.0079}_{-0.0079}$ & $-0.0029^{+0.0024}_{-0.0024}$ & $0.0055^{+0.0013}_{-0.0013}$ & $-0.0014^{+0.0004}_{-0.0004}$ \\
        \hline
                                    & $1 \leq \mu<4$    & $-0.226^{+0.267}_{-0.265}$ & $0.179^{+0.081}_{-0.080}$ & $0.0288^{+0.0061}_{-0.0061}$ & $-0.0036^{+0.0018}_{-0.0018}$ & $0.0058^{+0.0011}_{-0.0011}$ & $-0.0016^{+0.0003}_{-0.0003}$ \\
        $\geq 10.3$                 & $4 \leq \mu<10$   & $0.068^{+0.089}_{-0.090}$ & $0.033^{+0.038}_{-0.038}$ & $0.0334^{+0.0087}_{-0.0086}$ & $-0.0034^{+0.0025}_{-0.0025}$ & $0.0063^{+0.0015}_{-0.0015}$ & $-0.0014^{+0.0004}_{-0.0004}$ \\
                                    & $1 \leq \mu<10$   & $0.008^{+0.010}_{-0.010}$ & $0.077^{+0.043}_{-0.043}$ & $0.0289^{+0.0067}_{-0.0067}$ & $-0.0026^{+0.0020}_{-0.0021}$ & $0.0052^{+0.0011}_{-0.0011}$ & $-0.0013^{+0.0003}_{-0.0003}$ \\
        \hline
                                    & $1 \leq \mu<4$    & $-0.103^{+0.133}_{-0.134}$ & $0.113^{+0.050}_{-0.050}$ & $0.0273^{+0.0055}_{-0.0054}$ & $-0.0031^{+0.0017}_{-0.0017}$ & $0.0050^{+0.0009}_{-0.0009}$ & $-0.0013^{+0.0003}_{-0.0003}$  \\
        $\geq 10.5$                 & $4 \leq \mu<10$   & $0.043^{+0.058}_{-0.058}$ & $0.047^{+0.043}_{-0.043}$ & $0.0363^{+0.0090}_{-0.0090}$ & $-0.0048^{+0.0026}_{-0.0026}$ & $0.0062^{+0.0014}_{-0.0014}$ & $-0.0014^{+0.0004}_{-0.0004}$  \\
                                    & $1 \leq \mu<10$   & $-0.042^{+0.056}_{-0.057}$ & $0.087^{+0.040}_{-0.040}$ & $0.0307^{+0.0064}_{-0.0064}$ & $-0.0037^{+0.0019}_{-0.0019}$ & $0.0051^{+0.0011}_{-0.0011}$ & $-0.0012^{+0.0003}_{-0.0003}$  \\
        \hline
                                    & $1 \leq \mu<4$    & $-0.097^{+0.125}_{-0.126}$ & $0.096^{+0.041}_{-0.042}$ & $0.0280^{+0.0051}_{-0.0050}$ & $-0.0039^{+0.0015}_{-0.0015}$ & $0.0045^{+0.0008}_{-0.0008}$ & $-0.0012^{+0.0002}_{-0.0002}$\\
        $\geq 10.8$                 & $4 \leq \mu<10$   & $0.239^{+0.158}_{-0.156}$ & $0.023^{+0.030}_{-0.030}$ & $0.0330^{+0.0106}_{-0.0106}$ & $-0.0049^{+0.0032}_{-0.0033}$ & $0.0045^{+0.0016}_{-0.0016}$ & $-0.0011^{+0.0005}_{-0.0005}$ \\
                                    & $1 \leq \mu<10$   & $0.085^{+0.099}_{-0.099}$ & $0.020^{+0.025}_{-0.025}$ & $0.0295^{+0.0059}_{-0.0060}$ & $-0.0036^{+0.0017}_{-0.0017}$ & $0.0042^{+0.0009}_{-0.0009}$ & $-0.0009^{+0.0003}_{-0.0003}$ \\
        \hline
                                    & $1 \leq \mu<4$    & $-0.267^{+0.275}_{-0.277}$ & $0.180^{+0.083}_{-0.084}$ & $0.0312^{+0.0052}_{-0.0051}$ & $-0.0059^{+0.0015}_{-0.0016}$ & $0.0038^{+0.0009}_{-0.0009}$ & $-0.0011^{+0.0003}_{-0.0003}$ \\
        $\geq 11.0$                 & $4 \leq \mu<10$   & $0.449^{+0.169}_{-0.170}$ & $-0.040^{+0.049}_{-0.050}$ & $0.0453^{+0.0083}_{-0.0075}$ & $-0.0088^{+0.0026}_{-0.0020}$ & $-0.0005^{+0.0005}_{-0.0005}$ & $0.0003^{+0.0001}_{-0.0002}$ \\
                                    & $1 \leq \mu<10$   & $0.133^{+0.093}_{-0.095}$ & $0.0004^{+0.0006}_{-0.0006}$ & $0.0276^{+0.0054}_{-0.0051}$ & $-0.0032^{+0.0014}_{-0.0015}$ & $0.0031^{+0.0007}_{-0.0008}$ & $-0.0007^{+0.0002}_{-0.0002}$\\
        \hline
    \end{tabular}
\end{table*}

\begin{figure*}
	\includegraphics[width=\textwidth,height=\textheight,keepaspectratio]{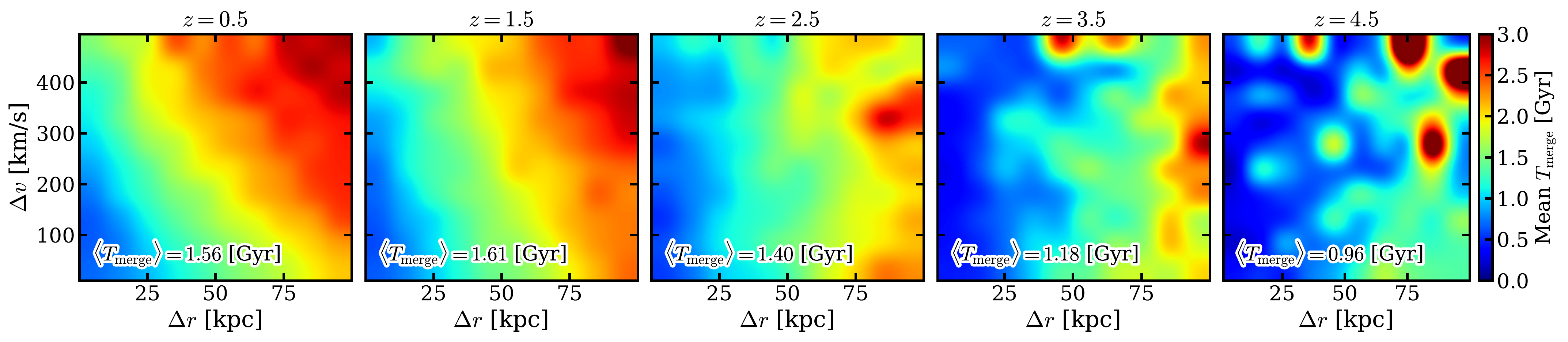}
	\caption{The mean merging timescales for pairs with $\log_{10}(m_1) \geq 10.3$ with $1 \leq \mu < 4$. Each panel includes pairs where the observed redshift of the primary galaxy falls within $\pm0.5$ of the noted central redshift.}
	\label{fig:mean_T}
\end{figure*}
The observation timescales shown in \citet{Oleary2021} are derived by mapping the pair fraction directly onto the merger rate without making any consideration over which pairs actually ended up merging. There we found that $\tobs\propto (1+z)^{-1}$ provided a reasonable translation from pair fractions to intrinsic galaxy merger rates. This is a weaker scaling than the $\tobs \propto (1+z)^{-2}$ proposed by \citet{Snyder2017}, and the $\tobs \propto H(z)^{-1/3}$ scaling suggested by \citet{Jiang2014}.

In this work, we identify individual pairs and track them until the point they finally merge. Here we should highlight that the value that we are measuring is the \textit{merging} timescale not the \textit{observation} timescale. The observation timescale tracks how long a pair would remain in the aperture set by the observer. The observation timescale should in principle also incorporate information on how long a pair remains in the mass and mass ratio criteria that has been set. Because \tobs is dependent on the particular observation, we have elected to fit to the merging timescales as it can be more readily applied to a broader range of selection criteria.

Figure~\ref{fig:mean_T} shows the mean merging timescale as a function of \dr and \dv in several redshift bins for our reference case. For $z\gtrsim0.5$ the data indicates merging timescales that decrease with redshift, as expected from previous results. We also find that the dependence on \dv has a stronger redshift scaling that the dependence on \dr. We found that a flat plane was sufficient to reproduce the data across a wide range of redshifts:
\begin{align}
    T_{\mathrm{merge}}(z,\dr,\dv) & = a + (b \dr) + (c \dv) \label{eqn:T} \\
    a & = a_0 + (1+z)a_z \nonumber\\
    b & = b_0 + (1+z)b_z \nonumber \\
    c & = c_0 + (1+z)c_z \nonumber 
\end{align}
In this formulation allowing each parameter to scale linearly with redshift provided the best reproduction of merger rates.

Table~\ref{tab:T_fit} Shows the best fit parameters from eq.~\ref{eqn:T} for a range of the same mass ranges shown in Table~\ref{tab:W_fit}. Generally eq.~\ref{eqn:T} provides an accurate description of the data for $1\lesssim z \lesssim 3.5$ for most mass ranges and mass ratios tested. Within this redshift range merging timescales increase with increasing \dv as seen in the data. However, outside this range the scaling with respect to $\dv$ undergoes a sign inversion due to the linear redshift scaling of the parameter $c$. This inversion results in a considerable under prediction of the merging timescales for small radii and $\dv\gtrsim200\kms$. However, since most of the major pairs sit below this \dv and at a more moderate \dr, we find this limited fit to the data does not strongly impact our reproduced major merger rates (see Section~\ref{sec:rates}).

We tested a variety of fitting functions and parameter redshift scaling to fit the pair merging timescales. Notably, allowing for $c\propto (1+z)^{c_z}$ alleviates the severity of under-predicted merging timescales for high \dv pairs. However this improved fit at high-$z$ comes at the expense of a significantly worse reproduction of merger rates for $z\lesssim 3$. The \dv slope scaling with redshift is the dominate parameter determining the goodness of fit for this parameterisation. The scaling with \dr is relatively stable with redshift and a suitable merger rate reproduction is possible if $b$ is kept static, though in that scenario we find a stronger over prediction of the merging timescales towards low-$z$, resulting in a more pronounced under prediction of the merger rate. Just as in \citet{Jiang2014} we express merging timescales that scale linearly in \dr, however in their work they elect for a physically motivated redshift scaling with $b\propto H(z)^{-1/3}$. When measuring average merging timescales for fixed $\dr_{\mathrm{max}}$ we find that our results are statistically consistent with that scaling. In practice we found this formulation did not offer an improved fit to the data, or reproduction of merger rates. If we allow a more free scaling where $b\propto H(z)^{\alpha}$ there is a slight improvement over $H(z)^{-1/3}$. Although, allowing that free scaling in $H(z)$ makes the physical interpretation of that term ambiguous, so we opt for the more simple linear redshift scaling, which produces the the better reproduction of merger rates.

In the next section we will combine these fits for merging probabilities and timescales and recover the merger rate of galaxies under a range of pair selection criteria.
\section{Reconstructing merger rates}
\label{sec:rates}
\begin{figure*}
	\includegraphics[width=\textwidth,height=\textheight,keepaspectratio]{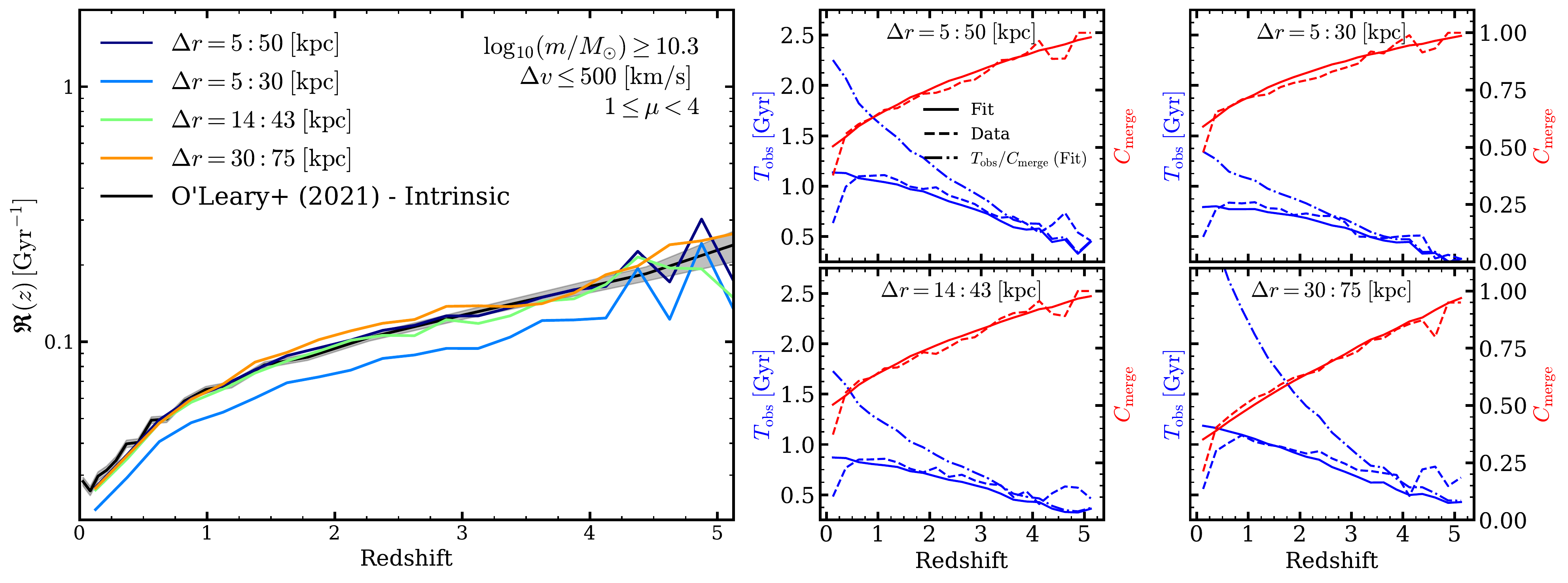}
	\caption{Galaxy merger rates reconstructed from mock pairs, the fitting functions of eq.~\ref{eqn:C} and eq.~\ref{eqn:T}. The leftmost panel shows the reconstructed rates under various apertures (coloured lines) against the intrinsic merger rates derived from trees in \citet{Oleary2021} (black lines). The right panels show how well our fitting functions (solid lines) reproduce the observation timescales (blue lines) and merging probabilities (red lines) seen in the data for the noted apertures (dashed lines). We also include the \textit{effective} observation timescales (blue dash-dot lines) defined as $T_{\mathrm{obs}}^{\mathrm{eff}}\equiv \tobs (z)/\cmerge (z)$.}
	\label{fig:rates}
\end{figure*}
\begin{figure}
	\includegraphics[width=\columnwidth]{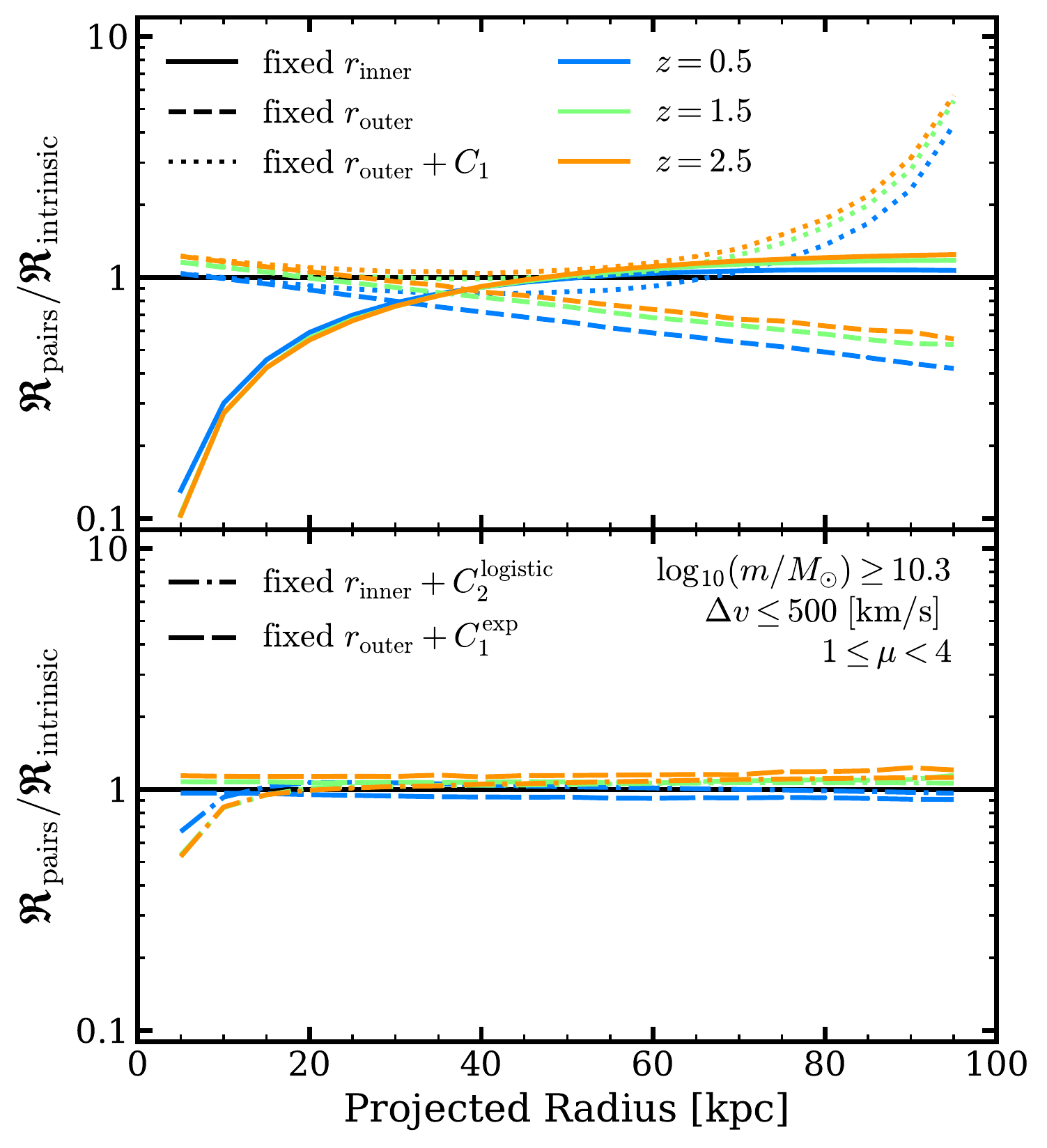}
\caption{The ratio of predicted to intrinsic merger rates as a function of radius, with various assumptions for area correction. In both panels, lines labelled ``fixed \rinner'' indicate the rate ratio assuming fixed inner aperture $\rinner=0\kpc$ and a variable outer aperture \router. Similarly, lines labelled ``fixed \router'' indicate the rate ratio assuming fixed outer aperture $\router=100\kpc$ and a variable inner aperture \rinner. Line colour indicates the redshift at which the ratio is taken. Values near $1$ indicate better agreement with intrinsic values. \textit{Top panel:} Solid and dashed lines show the rate ratio with no area corrections applied. The dotted line shows the rate ratio with the standard inner area correction of eq.~\ref{eqn:C1}. \textit{Bottom panel:} Dash-dot lines show prediction accuracy with the additionally correction factor shown in eq.~\ref{eqn:C2} which addresses incompleteness for limited \router. The long-dash lines show prediction accuracy using a newly implemented inner area correction eq.~\ref{eqn:C1B}.}
	\label{fig:ratio}
\end{figure}
Now that we have a handle on the merger probability, and the merging timescales for individual pairs we can apply these to our mock observations and recover the intrinsic merger rates. If we blindly use \tmerge in-place of \tobs the resulting merger rates we produce will be lower than those predicted in \citet{Oleary2021}. First we need to approximate the observation timescale by estimating how long each of our pairs reside in the observable aperture. Taking inspiration from the orphan position formula (eq.~\ref{eqn:radius}) we get:
\begin{equation}
    \label{eqn:tobs}
    \tobs = \tmerge \left[ 1 - \left(\frac{r_{\mathrm{inner}}}{\dr}\right)^{2}\right]\; ,
\end{equation}
where $r_{\mathrm{inner}}$ is the inner radius of the observable aperture. Additionally, we find it helpful to include an area correction to account for pairs that sit below the chosen aperture. We use the same correction shown in \citet{Ventou2019}:
\begin{equation}
    \label{eqn:C1}
    C_{1} = \frac{r_{\mathrm{inner}}^2}{r_{\mathrm{outer}}^2 - r_{\mathrm{inner}}^2}
\end{equation}

Incorporating our weighting scheme (eq.~\ref{eqn:C}), individualised observation timescales (eq.~\ref{eqn:tobs}) and the area correction (eq.~\ref{eqn:C1}), the merger rate formula from eq.~\ref{eqn:rate1} can be rewritten as;
\begin{equation}
    \label{eqn:R}
    \mathfrak{R}_{\mathrm{merge}} = C_{1} f_p \frac{\sum_{i}^{N_{p}}W(\dr, \dv, z)}{\sum_{i}^{N_{p}} \tobs(\dr, \dv, z)}
\end{equation}
Where $f_p$ is the pair fraction and $N_p$ is the number of pairs in the sample.

Figure~\ref{fig:rates} compares the galaxy merger rate derived from eq.~\ref{eqn:R} (coloured lines) with the merger rates shown in \citet{Oleary2021} (black line). In general the results from this are in excellent agreement with intrinsic rates. We find that we are consistently able to reproduce major merger rates to at least $z=4$ for a range of projected separations. We find a notable exception in the often used $\dr=5:30\kpc$ aperture \citep{Mundy2017, Duncan2019}, which under predicts the intrinsic merger rate at nearly every redshift. If we view the corresponding (upper right) panel of Figure~\ref{fig:rates} we can see that the fitting functions reproduce the observation timescales and merging probabilities of the data just as well as any of the other apertures we tested. Taken in the context of Figure~\ref{fig:pair_dist} and absent any other effects that might influence the observation timescales for this selection criteria we can deduce that this aperture is ill-suited for reconstructing the underlying merger rate as the small outer radius excludes a significant fraction of pairs undergoing a merger. The under prediction exhibited by this particular selection criteria is present for all major merger rates that we tested. 

In Figure~\ref{fig:ratio} we show the accuracy of our predictions, exhibited as the ratio of predicted to intrinsic merger rate, as a function of observable aperture at three redshifts. For lines labelled ``fixed \rinner'' we set a constant $\rinner=0\kpc$ while increasing \router out to $100\kpc$. Lines labelled ``fixed \router'' hold a fixed out radius at $\router=100\kpc$ with a variable inner radius \rinner. In the top panel, solid and dashed lines show the accuracy of our model absent any area corrections. For ``fixed \rinner", solid lines, we can see clearly if $\router \lesssim 50\kpc$ our predictions would undercut the intrinsic merger rate. This coincides directly with the under prediction shown in the $\dr=5:30\kpc$ range of Figure~\ref{fig:rates}, and illustrates the necessity for an additional corrective factor. To address the pair incompleteness due to limited outer aperture we introduce the following correction: 
\begin{equation}
    C_{2}^{\mathrm{logistic}} = \left[\frac{L}{1+\exp(-k r_{\mathrm{outer}})} -1\right]^{-1}\;,
    \label{eqn:C2}
\end{equation}
where $L$ and $k$ are free parameters. This functions once again takes the form of a (half) logistic curve and is fit to the solid lines in the top panel of Figure~\ref{fig:ratio}. When fitting we assume $z=1.0$, $\log(m/M_{\odot})\geq 10.3$, and $1\leq\mu<4$. Under these conditions, and a simple non-linear least squares fit, we find $L=2.13\pm0.02$ and $k=0.050\pm0.002\kpc^{-1}$. The bottom panel illustrates the impact of this new correction, dash-dotted lines. Here we can see that this correction substantially improves the accuracy of predicted results, particularly for $\router \lesssim 50\kpc$. We found these best fit parameters are suitable to correct major merger rates for $\log(m/M_{\odot})\geq 9.5$ and $\log(m/M_{\odot})\geq 11.0$, indicating minimal mass dependency.

Additionally, The top panel of Figure~\ref{fig:ratio} shows that that eq.~\ref{eqn:C1} provides a reasonable correction for non-optimal \router (dotted lines), but predictably deviates as $\rinner\rightarrow\router$. The corresponding uncorrected result, dashed lines, suggest an exponential correction maybe be more appropriate. Thus we introduce:
\begin{equation}
    C_{1}^{\exp} = \alpha \exp\left(\frac{r_{\mathrm{inner}}}{\beta}\right)\;,
    \label{eqn:C1B}
\end{equation}
where $\alpha$ and $\beta$ are free parameters. Using the same fitting criteria noted above we find $\alpha=0.889\pm0.003$ and $\beta=106\pm1\kpc$. In the lower panel of Figure~\ref{fig:ratio} (long-dash lines) we can see this updated correction factor improves the accuracy of predicted rates even when \rinner is similar to \router. Just as before we found this correction factor broadly applicable to major merger rates without the need to refit for higher or lower mass cuts. However, in practice this updated formulation does not provide significantly improved results compared to eq.~\ref{eqn:C1} when applied to commonly used apertures.

Further for the lowest mass bin we fit, $9.0 \leq \log_{10}(m_1/M_{\odot}) < 10.0$, we found our fitting functions fail to accurately reproduce the underlying major merger rate at any redshift. Here we generally find a consistent over prediction in merger rates by a factor of $\sim3$ to $\sim5$. We nonetheless include fits to this mass range for completeness.

When considering minor mergers we find that rates can be well reproduced out to $z\approx3.5$ for all mass ranges tested. However, due to the poorer performance of the fitting function at high $z$ and \dv, we find a significant under prediction of the merging timescales. This translates into an over prediction of the merger rate by a factor $\sim1.5$ near $z\approx3.5$ to a factor $\sim5$ near $z\approx5$. This mismatch is most pronounced where $\log_{10}(m_{1}/M_{\odot})\geq 10.0$.

\begin{figure}
	\includegraphics[width=\columnwidth]{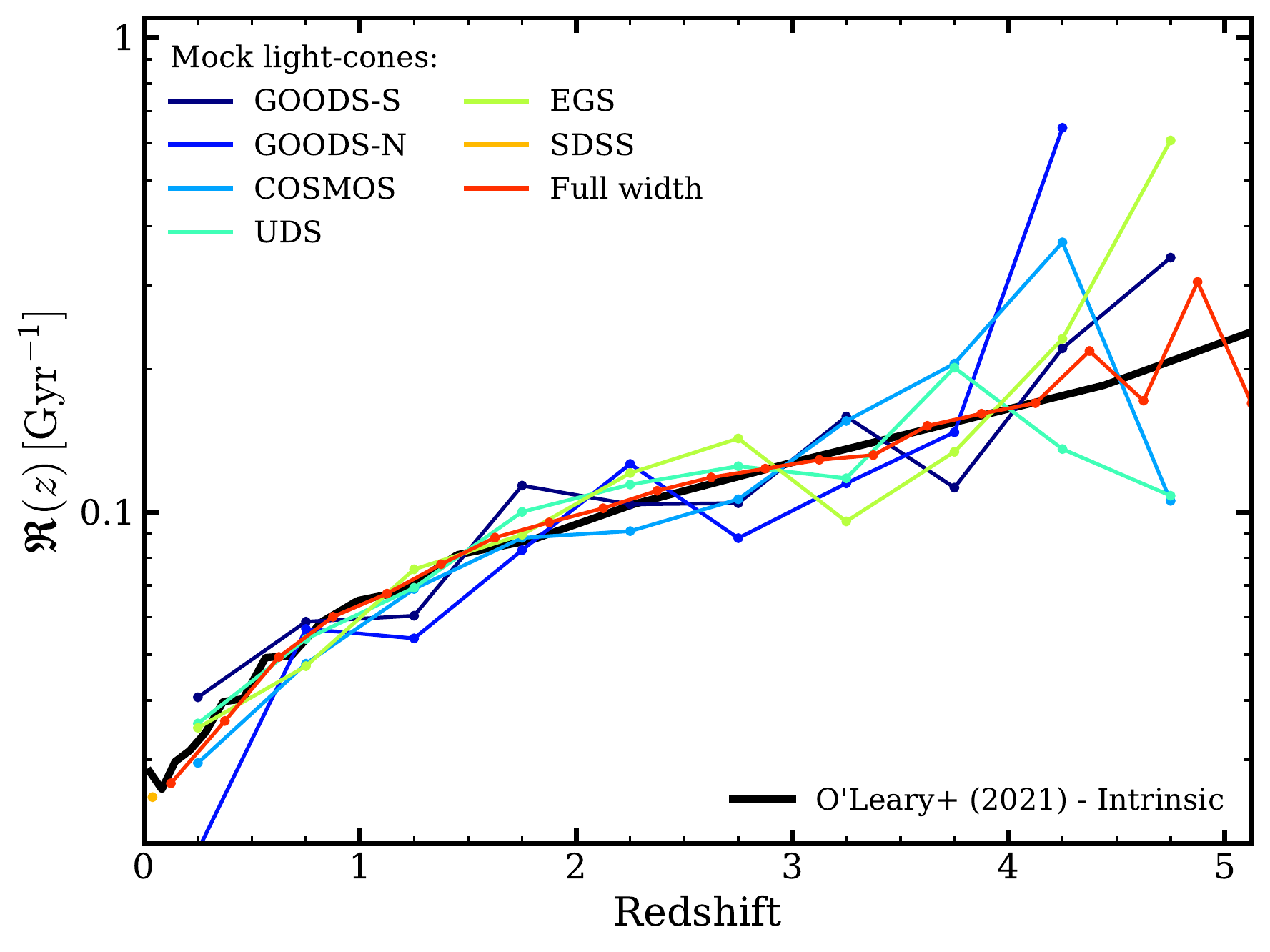}
\caption{The merger rates determined from mock light-cones. Rates are computed for $\log_{10}(m_{1}/M_{\odot}) \geq 10.3$,  $5\leq\dr<50\kpc$ projected separation, $\dv \leq 500\kms$ and $1\leq\mu<4$. The solid black line indicates intrinsic merger rates as shown in \citet{Oleary2021}. Coloured lines show merger rates determined from mock light-cones using eq~\ref{eqn:R}. Here we do not place any redshift restraints on the light-cone catalogues that would more closely resemble the observables limits of the noted surveys.}
	\label{fig:cone_rates}
\end{figure}

Figure~\ref{fig:cone_rates} illustrates the derived merger rates from the mock light-cone catalogues described in Section~\ref{sec:lightcone}. Here we can see that our formulae show excellent reproduction of underlying merger rates for a range of cone geometry, where each sample contains a unique set of pairs.

\section{Discussion}
\label{sec:conclusions}
\subsection{Fitting at low and high redshift}
Although we are able to confidently reproduce merger rates using our fitting functions there are some regions where care should be taken in how results can be interpreted. Notably, the data indicates that for $z\lesssim 1$ the merger probability tends to zero. This trend makes sense recalling that under our definition of merger probability, that a pair observed at zero should have zero per cent chance to merge by zero. Looking back to the merging timescales, here we also note that the merging timescale turns over and begins to sharply decrease for $z\lesssim 1$. This turn over does not indicate a physical process that suppresses the merging timescales for low $z$. This data is constructed only from pairs that \textit{did} merge by $z=0$ so must necessarily decrease to reflect the decreasing time remaining for a pair to merge.

With this data we are left to decide whether we should fit to this low redshift regime given the bias towards pairs that merged quickly. In selecting fitting functions, we also tested variations that reproduced the trend to zero for both merging probabilities and timescales. Those fitting functions that worked for that low redshift region typically struggled more to fit the data, and reproduce  merger rates at intermediate and high redshift. For this reason we opted for fitting functions that over shoot the data at low-$z$ as they showed better performance when reconstructing merger rates.

\subsection{Completeness}
In Section~\ref{sec:rates} we illustrated how improper selection of observable aperture can result in large under predictions in merger rates due to missing pairs. Although corrections already exist to address incompleteness due to large \rinner, there are currently no widely used treatments to counteract incompleteness due to small \router. In this section we provided updated correction functions that successfully improve the accuracy of predicted results for a large range of observable area, and mass cuts.

The primary goal of this work is to place better constraints on \tobs and \cmerge, not address other observable restrictions that might impact pair counts or their translation to rates. Therefore, for these completeness corrections we only tested their application to major mergers $1\leq\mu<4$ for select mass ranges. Although we do not see evidence for any strong mass scaling in the function parameters, future work should explore the mass and mass ratio dependencies in greater detail. Additionally, when fitting these functions we elected to fit only at a single redshift, $z=1.0$. The results shown in Figure~\ref{fig:ratio} suggest either of these functions might have a weak redshift dependence. Future work with these functions would benefit from a fitting routine that considers the entire redshift range of analysis, or even an additional parameterisation for redshift dependency.

\subsection{Sources of uncertainty}
The results shown here stand as an extension to the results of \citet{Oleary2021}, what we show here does not explore the complete dependencies of these formulations on our model assumptions. Thus, there remain several sources of uncertainty that could impact these results that we do not quantify in this work. 

We have seen that merging timescales, and merging probabilities are strongly dependent on the relative line of sight velocities between each galaxy in the pair. At present \textsc{emerge} has no prescription for updating velocities of orphan galaxies. Currently, orphans simply inherit the velocity of their last resolved halo. Particularly at low masses, where orphans make up a large portion of the galaxy stellar mass function, this may alter the distribution of velocities in observed pairs, as well as the assumed fitted parameters to eq.~\ref{eqn:C} and eq.~\ref{eqn:T}. For the example case we show in this work, major pairs consisting of at least 1 orphan galaxy constitute as much as $\sim30$ per cent of all pairs at $z\approx0$ falling to around $\sim5$ per cent by $z\approx4$. For the lowest mass bin shown in Table~\ref{tab:W_fit} and Table~\ref{tab:T_fit} the orphan pair fraction increases to around $\sim60$ per cent by $z\approx0$. This may be one of the sources for the poor reproduction of major merger rates in this mass range.

Additionally, merging in \textsc{emerge} is entirely defined by the dynamical friction formula chosen. In practice there are other aspects of the physical system that should be considered. In this model a satellite galaxy can be placed arbitrarily close to its host system but will only be merged at $t_{\mathrm{df}}$. In practice these satellites may be affected by the radial extent of the host system resulting in a galaxy merging sooner than $t_{\mathrm{df}}$. Such mechanisms could be a driver in the lower galaxy merger rates exhibited by our model compared with others. If merging timescales are artificially long, this would similarly extend the assumed observation timescales of satellite galaxies under our current formulation. Further, including mechanisms that reduce merging timescales may also impact the distribution of observed pairs, our chosen fitting functions may not be suitable under such model variations.

These results would benefit from a more complete study on how model assumptions drive pair fractions and merging timescales. Our fitting functions assume a narrow description of the observation timescale. In this work we assume the observation timescale is driven entirely by the time a galaxy pair fulfils the \dr criteria chosen, which linked directly to the average merging timescale of such a pair. This notably neglects the star formation properties in each pair. For $z\gtrsim2$ high star formation rates result in a strong mass evolution in the galaxy population. This should reduce the amount of time any given pair can satisfying the mass and mass ratio criteria. In this work we largely reference pair selections set by a lower stellar mass cut, which mitigates the impact on observation timescales by pairs moving out of the noted mass bin. However it places no constraints on the impact from pairs that evolve outside of the mass ratio criteria. Under the broad assumption that these mechanisms would only serve to reduce the observation timescale, our results indicate that the observation timescale is instead largely driven by the time spent in the chosen aperture, as our fitting functions generally do not \textit{under} predict the merger rate.

\section{Conclusions}

In this work we show that with our empirical model \textsc{emerge} model we are able to construct fitting functions for galaxy merging probabilities and timescales that can accurately recover the intrinsic merger rate across range of selection criteria. To that end we provide best fit model parameters for a wide range of commonly used stellar mass intervals, and mass ratios. 

For a given aperture, the data indicates \tobs is approximately linear between $z\sim1$ and $z\sim4$, but $T_{\mathrm{obs}}^{\mathrm{eff}}$ can range between a linear scaling and power-law scaling depending on the aperture chosen. However, we reinforce a \tobs that does not evolve as strongly as \citet{Snyder2017} or \citet{Jiang2014}. We can conclude from our results that pairs undergoing a major merger have a $\tobs$ that is primarily driven by dynamical processes to at least $\sim3.5$.

We further show that it is not necessary to fit the observation timescales directly, but instead fit to merging timescales and derive the selection criteria dependent observation timescales using a formula that approximates radial decay in haloes. Additionally, we can show with these methods that not all pair selection criteria are equally suited to determining merger rates. If the outer radius of the observable aperture is too small, a non-negligible fraction of merging systems can be missed, resulting in an under prediction of the galaxy merger rate. To combat this outer area incompleteness we suggest a new correction factor that should be applied to observations with sub-optimal apertures.

Finally, in the last section we discussed areas where these results could be improved with further study. At high-$z$ ($\gtrsim4$) box size limitations prevent us from placing tighter constraints on merger timescale evolution of massive objects due to low number counts. At the lowest masses our fitting functions are inadequate to reliably reproduce the underlying galaxy merger rate, at these masses and at high redshift \tobs may be dominated by high start formation rates, which reduce the time a pair spends in the mass and mass ratio bin.
\section*{Acknowledgements}

We thank all authors who provide their data in electronic form. We are also grateful to Ulrich Steinwandel, Ludwig Böss, Rhea-Silvia Rhemus, Tadziu Hoffmann, Klaus Dolag and Julien Wolf for enlightening discussions. The cosmological simulations used in this work were carried out at the Odin Cluster at the Max Planck Computing and Data Facility in Garching. BPM and JAO acknowledges an Emmy Noether grant funded by the Deutsche Forschungsgemeinschaft (DFG, German Research Foundation) -- MO 2979/1-1. Finally, we thank the developers of \texttt{Astropy} \citep{Astropy-Collaboration:2013aa,Astropy-Collaboration:2018aa}, \texttt{NumPy} \citep{van-der-Walt:2011aa}, \texttt{SciPy} \citep{Virtanen:2020aa}, \texttt{Jupyter} \citep{Ragan-Kelley:2014aa}, \texttt{Matplotlib} \citep{Hunter:2007aa}, \texttt{HaloTools} \citep{Hearin:2017aa} for their very useful free software. The Astrophysics Data Service (ADS) and \texttt{arXiv} preprint repository were used extensively in this work.

\section*{Data availability}
The source code of \textsc{emerge} and sample galaxy merger trees are available on \url{https://github.com/bmoster/emerge}. The derived data and analysis scripts used in this article can be found at \url{https://github.com/jaoleary}.




\bibliographystyle{mnras}
\bibliography{references} 

\begin{thebibliography}{}
\makeatletter
\relax
\def\mn@urlcharsother{\let\do\@makeother \do\$\do\&\do\#\do\^\do\_\do\%\do\~}
\def\mn@doi{\begingroup\mn@urlcharsother \@ifnextchar [ {\mn@doi@}
  {\mn@doi@[]}}
\def\mn@doi@[#1]#2{\def\@tempa{#1}\ifx\@tempa\@empty \href
  {http://dx.doi.org/#2} {doi:#2}\else \href {http://dx.doi.org/#2} {#1}\fi
  \endgroup}
\def\mn@eprint#1#2{\mn@eprint@#1:#2::\@nil}
\def\mn@eprint@arXiv#1{\href {http://arxiv.org/abs/#1} {{\tt arXiv:#1}}}
\def\mn@eprint@dblp#1{\href {http://dblp.uni-trier.de/rec/bibtex/#1.xml}
  {dblp:#1}}
\def\mn@eprint@#1:#2:#3:#4\@nil{\def\@tempa {#1}\def\@tempb {#2}\def\@tempc
  {#3}\ifx \@tempc \@empty \let \@tempc \@tempb \let \@tempb \@tempa \fi \ifx
  \@tempb \@empty \def\@tempb {arXiv}\fi \@ifundefined
  {mn@eprint@\@tempb}{\@tempb:\@tempc}{\expandafter \expandafter \csname
  mn@eprint@\@tempb\endcsname \expandafter{\@tempc}}}

\bibitem[\protect\citeauthoryear{{Abraham}, {van den Bergh}  \&
  {Nair}}{{Abraham} et~al.}{2003}]{Abraham2003}
{Abraham} R.~G.,  {van den Bergh} S.,   {Nair} P.,  2003, \mn@doi [\apj]
  {10.1086/373919}, \href
  {https://ui.adsabs.harvard.edu/abs/2003ApJ...588..218A} {588, 218}

\bibitem[\protect\citeauthoryear{{Abruzzo}, {Narayanan}, {Dav{\'e}}  \&
  {Thompson}}{{Abruzzo} et~al.}{2018}]{Abruzzo2018}
{Abruzzo} M.~W.,  {Narayanan} D.,  {Dav{\'e}} R.,   {Thompson} R.,  2018, arXiv
  e-prints, \href {https://ui.adsabs.harvard.edu/abs/2018arXiv180302374A} {p.
  arXiv:1803.02374}

\bibitem[\protect\citeauthoryear{{Astropy Collaboration} et~al.,}{{Astropy
  Collaboration} et~al.}{2013}]{Astropy-Collaboration:2013aa}
{Astropy Collaboration} et~al., 2013, \mn@doi [\aap]
  {10.1051/0004-6361/201322068}, \href
  {https://ui.adsabs.harvard.edu/abs/2013A&A...558A..33A} {558, A33}

\bibitem[\protect\citeauthoryear{{Astropy Collaboration} et~al.,}{{Astropy
  Collaboration} et~al.}{2018}]{Astropy-Collaboration:2018aa}
{Astropy Collaboration} et~al., 2018, \mn@doi [\aj] {10.3847/1538-3881/aabc4f},
  \href {https://ui.adsabs.harvard.edu/abs/2018AJ....156..123A} {156, 123}

\bibitem[\protect\citeauthoryear{{Behroozi}, {Wechsler}  \& {Wu}}{{Behroozi}
  et~al.}{2013a}]{rockstar}
{Behroozi} P.~S.,  {Wechsler} R.~H.,   {Wu} H.-Y.,  2013a, \mn@doi [\apj]
  {10.1088/0004-637X/762/2/109}, \href
  {http://adsabs.harvard.edu/abs/2013ApJ...762..109B} {762, 109}

\bibitem[\protect\citeauthoryear{{Behroozi}, {Wechsler}, {Wu}, {Busha},
  {Klypin}  \& {Primack}}{{Behroozi} et~al.}{2013b}]{ctrees}
{Behroozi} P.~S.,  {Wechsler} R.~H.,  {Wu} H.-Y.,  {Busha} M.~T.,  {Klypin}
  A.~A.,   {Primack} J.~R.,  2013b, \mn@doi [\apj]
  {10.1088/0004-637X/763/1/18}, \href
  {http://adsabs.harvard.edu/abs/2013ApJ...763...18B} {763, 18}

\bibitem[\protect\citeauthoryear{{Behroozi}, {Wechsler}  \&
  {Conroy}}{{Behroozi} et~al.}{2013c}]{Behroozi2013d}
{Behroozi} P.~S.,  {Wechsler} R.~H.,   {Conroy} C.,  2013c, \mn@doi [\apj]
  {10.1088/0004-637X/770/1/57}, \href
  {https://ui.adsabs.harvard.edu/abs/2013ApJ...770...57B} {770, 57}

\bibitem[\protect\citeauthoryear{{Behroozi}, {Wechsler}, {Hearin}  \&
  {Conroy}}{{Behroozi} et~al.}{2019}]{Behroozi2019}
{Behroozi} P.,  {Wechsler} R.~H.,  {Hearin} A.~P.,   {Conroy} C.,  2019,
  \mn@doi [\mnras] {10.1093/mnras/stz1182}, \href
  {https://ui.adsabs.harvard.edu/abs/2019MNRAS.488.3143B} {488, 3143}

\bibitem[\protect\citeauthoryear{Benson}{Benson}{2012}]{Benson2012}
Benson A.~J.,  2012, \mn@doi [New Astronomy]
  {https://doi.org/10.1016/j.newast.2011.07.004}, 17, 175

\bibitem[\protect\citeauthoryear{{Binney} \& {Tremaine}}{{Binney} \&
  {Tremaine}}{1987}]{Binney87}
{Binney} J.,  {Tremaine} S.,  1987, {Galactic dynamics}.
{Pinceton University Press}, {Princeton University, Princeton, NJ}

\bibitem[\protect\citeauthoryear{{Bluck} et~al.,}{{Bluck}
  et~al.}{2019}]{Bluck2019}
{Bluck} A. F.~L.,  et~al., 2019, \mn@doi [\mnras] {10.1093/mnras/stz363}, \href
  {https://ui.adsabs.harvard.edu/abs/2019MNRAS.485..666B} {485, 666}

\bibitem[\protect\citeauthoryear{{Bois} et~al.,}{{Bois}
  et~al.}{2011}]{Bois2011}
{Bois} M.,  et~al., 2011, \mn@doi [\mnras] {10.1111/j.1365-2966.2011.19113.x},
  \href {https://ui.adsabs.harvard.edu/abs/2011MNRAS.416.1654B} {416, 1654}

\bibitem[\protect\citeauthoryear{{Bower}, {Benson}, {Malbon}, {Helly}, {Frenk},
  {Baugh}, {Cole}  \& {Lacey}}{{Bower} et~al.}{2006}]{Bower2006}
{Bower} R.~G.,  {Benson} A.~J.,  {Malbon} R.,  {Helly} J.~C.,  {Frenk} C.~S.,
  {Baugh} C.~M.,  {Cole} S.,   {Lacey} C.~G.,  2006, \mn@doi [\mnras]
  {10.1111/j.1365-2966.2006.10519.x}, \href
  {http://adsabs.harvard.edu/abs/2006MNRAS.370..645B} {370, 645}

\bibitem[\protect\citeauthoryear{{Boylan-Kolchin}, {Ma}  \&
  {Quataert}}{{Boylan-Kolchin} et~al.}{2008}]{Boylan-Kolchin2008}
{Boylan-Kolchin} M.,  {Ma} C.-P.,   {Quataert} E.,  2008, \mn@doi [\mnras]
  {10.1111/j.1365-2966.2007.12530.x}, \href
  {https://ui.adsabs.harvard.edu/\#abs/2008MNRAS.383...93B} {383, 93}

\bibitem[\protect\citeauthoryear{{Bundy}, {Fukugita}, {Ellis}, {Targett},
  {Belli}  \& {Kodama}}{{Bundy} et~al.}{2009}]{Bundy2009}
{Bundy} K.,  {Fukugita} M.,  {Ellis} R.~S.,  {Targett} T.~A.,  {Belli} S.,
  {Kodama} T.,  2009, \mn@doi [\apj] {10.1088/0004-637X/697/2/1369}, \href
  {https://ui.adsabs.harvard.edu/abs/2009ApJ...697.1369B} {697, 1369}

\bibitem[\protect\citeauthoryear{{Choi}, {Somerville}, {Ostriker}, {Naab}  \&
  {Hirschmann}}{{Choi} et~al.}{2018}]{Choi2018}
{Choi} E.,  {Somerville} R.~S.,  {Ostriker} J.~P.,  {Naab} T.,   {Hirschmann}
  M.,  2018, \mn@doi [\apj] {10.3847/1538-4357/aae076}, \href
  {https://ui.adsabs.harvard.edu/abs/2018ApJ...866...91C} {866, 91}

\bibitem[\protect\citeauthoryear{{Conroy} \& {Wechsler}}{{Conroy} \&
  {Wechsler}}{2009}]{Conroy2009}
{Conroy} C.,  {Wechsler} R.~H.,  2009, \mn@doi [\apj]
  {10.1088/0004-637X/696/1/620}, \href
  {https://ui.adsabs.harvard.edu/abs/2009ApJ...696..620C} {696, 620}

\bibitem[\protect\citeauthoryear{{Conselice}, {Bershady}, {Dickinson}  \&
  {Papovich}}{{Conselice} et~al.}{2003}]{Conselice2003}
{Conselice} C.~J.,  {Bershady} M.~A.,  {Dickinson} M.,   {Papovich} C.,  2003,
  \mn@doi [\aj] {10.1086/377318}, \href
  {https://ui.adsabs.harvard.edu/abs/2003AJ....126.1183C} {126, 1183}

\bibitem[\protect\citeauthoryear{{Dubois} et~al.,}{{Dubois}
  et~al.}{2014}]{Dubois2014}
{Dubois} Y.,  et~al., 2014, \mn@doi [\mnras] {10.1093/mnras/stu1227}, \href
  {https://ui.adsabs.harvard.edu/abs/2014MNRAS.444.1453D} {444, 1453}

\bibitem[\protect\citeauthoryear{{Duncan} et~al.,}{{Duncan}
  et~al.}{2019}]{Duncan2019}
{Duncan} K.,  et~al., 2019, \mn@doi [\apj] {10.3847/1538-4357/ab148a}, \href
  {https://ui.adsabs.harvard.edu/abs/2019ApJ...876..110D} {876, 110}

\bibitem[\protect\citeauthoryear{{Foreman-Mackey}, {Hogg}, {Lang}  \&
  {Goodman}}{{Foreman-Mackey} et~al.}{2013}]{Foreman-Mackey2013}
{Foreman-Mackey} D.,  {Hogg} D.~W.,  {Lang} D.,   {Goodman} J.,  2013, \mn@doi
  [\pasp] {10.1086/670067}, \href
  {https://ui.adsabs.harvard.edu/abs/2013PASP..125..306F} {125, 306}

\bibitem[\protect\citeauthoryear{{Gao} et~al.,}{{Gao} et~al.}{2020}]{Gao2020}
{Gao} F.,  et~al., 2020, \mn@doi [\aap] {10.1051/0004-6361/201937178}, \href
  {https://ui.adsabs.harvard.edu/abs/2020A&A...637A..94G} {637, A94}

\bibitem[\protect\citeauthoryear{{Goodman} \& {Weare}}{{Goodman} \&
  {Weare}}{2010}]{Goodman2010}
{Goodman} J.,  {Weare} J.,  2010, \mn@doi [CAMCoS] {10.2140/camcos.2010.5.65},
  \href {https://ui.adsabs.harvard.edu/abs/2010CAMCS...5...65G} {5, 65}

\bibitem[\protect\citeauthoryear{{Hahn} \& {Abel}}{{Hahn} \&
  {Abel}}{2011}]{music}
{Hahn} O.,  {Abel} T.,  2011, \mn@doi [\mnras]
  {10.1111/j.1365-2966.2011.18820.x}, \href
  {http://adsabs.harvard.edu/abs/2011MNRAS.415.2101H} {415, 2101}

\bibitem[\protect\citeauthoryear{{Hearin} et~al.,}{{Hearin}
  et~al.}{2017}]{Hearin:2017aa}
{Hearin} A.~P.,  et~al., 2017, \mn@doi [\aj] {10.3847/1538-3881/aa859f}, \href
  {https://ui.adsabs.harvard.edu/abs/2017AJ....154..190H} {154, 190}

\bibitem[\protect\citeauthoryear{{Henriques}, {White}, {Thomas}, {Angulo},
  {Guo}, {Lemson}, {Springel}  \& {Overzier}}{{Henriques}
  et~al.}{2015}]{Henriques2015}
{Henriques} B. M.~B.,  {White} S. D.~M.,  {Thomas} P.~A.,  {Angulo} R.,  {Guo}
  Q.,  {Lemson} G.,  {Springel} V.,   {Overzier} R.,  2015, \mn@doi [\mnras]
  {10.1093/mnras/stv705}, \href
  {https://ui.adsabs.harvard.edu/abs/2015MNRAS.451.2663H} {451, 2663}

\bibitem[\protect\citeauthoryear{{Hirschmann}, {Dolag}, {Saro}, {Bachmann},
  {Borgani}  \& {Burkert}}{{Hirschmann} et~al.}{2014}]{Hirschmann2014}
{Hirschmann} M.,  {Dolag} K.,  {Saro} A.,  {Bachmann} L.,  {Borgani} S.,
  {Burkert} A.,  2014, \mn@doi [\mnras] {10.1093/mnras/stu1023}, \href
  {https://ui.adsabs.harvard.edu/abs/2014MNRAS.442.2304H} {442, 2304}

\bibitem[\protect\citeauthoryear{{Hopkins} et~al.,}{{Hopkins}
  et~al.}{2018}]{Hopkins2018}
{Hopkins} P.~F.,  et~al., 2018, \mn@doi [\mnras] {10.1093/mnras/sty1690}, \href
  {https://ui.adsabs.harvard.edu/abs/2018MNRAS.480..800H} {480, 800}

\bibitem[\protect\citeauthoryear{{Hunter}}{{Hunter}}{2007}]{Hunter:2007aa}
{Hunter} J.~D.,  2007, \mn@doi [CiSE] {10.1109/MCSE.2007.55}, \href
  {https://ui.adsabs.harvard.edu/abs/2007CSE.....9...90H} {9, 90}

\bibitem[\protect\citeauthoryear{{Jesseit}, {Cappellari}, {Naab}, {Emsellem}
  \& {Burkert}}{{Jesseit} et~al.}{2009}]{Jesseit2009}
{Jesseit} R.,  {Cappellari} M.,  {Naab} T.,  {Emsellem} E.,   {Burkert} A.,
  2009, \mn@doi [\mnras] {10.1111/j.1365-2966.2009.14984.x}, \href
  {https://ui.adsabs.harvard.edu/abs/2009MNRAS.397.1202J} {397, 1202}

\bibitem[\protect\citeauthoryear{Jiang, Jing  \& Han}{Jiang
  et~al.}{2014}]{Jiang2014}
Jiang C.~Y.,  Jing Y.~P.,   Han J.,  2014, \mn@doi [\apj]
  {10.1088/0004-637x/790/1/7}, 790, 7

\bibitem[\protect\citeauthoryear{{Kampczyk} et~al.,}{{Kampczyk}
  et~al.}{2007}]{Kampczyk2007}
{Kampczyk} P.,  et~al., 2007, \mn@doi [\apjs] {10.1086/516594}, \href
  {https://ui.adsabs.harvard.edu/abs/2007ApJS..172..329K} {172, 329}

\bibitem[\protect\citeauthoryear{{Kartaltepe} et~al.,}{{Kartaltepe}
  et~al.}{2010}]{Kartaltepe2010}
{Kartaltepe} J.~S.,  et~al., 2010, \mn@doi [\apj] {10.1088/0004-637X/721/1/98},
  \href {https://ui.adsabs.harvard.edu/abs/2010ApJ...721...98K} {721, 98}

\bibitem[\protect\citeauthoryear{{Khalatyan}, {Cattaneo}, {Schramm},
  {Gottl{\"o}ber}, {Steinmetz}  \& {Wisotzki}}{{Khalatyan}
  et~al.}{2008}]{Khalatyan2008}
{Khalatyan} A.,  {Cattaneo} A.,  {Schramm} M.,  {Gottl{\"o}ber} S.,
  {Steinmetz} M.,   {Wisotzki} L.,  2008, \mn@doi [\mnras]
  {10.1111/j.1365-2966.2008.13093.x}, \href
  {https://ui.adsabs.harvard.edu/abs/2008MNRAS.387...13K} {387, 13}

\bibitem[\protect\citeauthoryear{{Kitzbichler} \& {White}}{{Kitzbichler} \&
  {White}}{2007}]{Kitzbichler2008}
{Kitzbichler} M.~G.,  {White} S.~D.~M.,  2007, \mn@doi [\mnras]
  {10.1111/j.1365-2966.2007.11458.x}, \href
  {https://ui.adsabs.harvard.edu/abs/2007MNRAS.376....2K} {376, 2}

\bibitem[\protect\citeauthoryear{{Koekemoer} et~al.,}{{Koekemoer}
  et~al.}{2011}]{Koekemoer2011}
{Koekemoer} A.~M.,  et~al., 2011, \mn@doi [\apjs] {10.1088/0067-0049/197/2/36},
  \href {https://ui.adsabs.harvard.edu/abs/2011ApJS..197...36K} {197, 36}

\bibitem[\protect\citeauthoryear{{Lewis}, {Challinor}  \& {Lasenby}}{{Lewis}
  et~al.}{2000}]{camb}
{Lewis} A.,  {Challinor} A.,   {Lasenby} A.,  2000, \mn@doi [\apj]
  {10.1086/309179}, \href
  {https://ui.adsabs.harvard.edu/abs/2000ApJ...538..473L} {538, 473}

\bibitem[\protect\citeauthoryear{{Lin} et~al.,}{{Lin} et~al.}{2008}]{Lin2008}
{Lin} L.,  et~al., 2008, \mn@doi [\apj] {10.1086/587928}, \href
  {https://ui.adsabs.harvard.edu/abs/2008ApJ...681..232L} {681, 232}

\bibitem[\protect\citeauthoryear{{L{\'o}pez-Sanjuan}, {Balcells},
  {P{\'e}rez-Gonz{\'a}lez}, {Barro}, {Garc{\'\i}a-Dab{\'o}}, {Gallego}  \&
  {Zamorano}}{{L{\'o}pez-Sanjuan} et~al.}{2009}]{Lopez-Sanjuan2009}
{L{\'o}pez-Sanjuan} C.,  {Balcells} M.,  {P{\'e}rez-Gonz{\'a}lez} P.~G.,
  {Barro} G.,  {Garc{\'\i}a-Dab{\'o}} C.~E.,  {Gallego} J.,   {Zamorano} J.,
  2009, \mn@doi [\aap] {10.1051/0004-6361/200911923}, \href
  {https://ui.adsabs.harvard.edu/abs/2009A&A...501..505L} {501, 505}

\bibitem[\protect\citeauthoryear{{Lotz}, {Jonsson}, {Cox}  \& {Primack}}{{Lotz}
  et~al.}{2008}]{Lotz2008}
{Lotz} J.~M.,  {Jonsson} P.,  {Cox} T.~J.,   {Primack} J.~R.,  2008, \mn@doi
  [\mnras] {10.1111/j.1365-2966.2008.14004.x}, \href
  {https://ui.adsabs.harvard.edu/abs/2008MNRAS.391.1137L} {391, 1137}

\bibitem[\protect\citeauthoryear{{Lotz}, {Jonsson}, {Cox}  \& {Primack}}{{Lotz}
  et~al.}{2010}]{Lotz2010}
{Lotz} J.~M.,  {Jonsson} P.,  {Cox} T.~J.,   {Primack} J.~R.,  2010, \mn@doi
  [\mnras] {10.1111/j.1365-2966.2010.16268.x}, \href
  {https://ui.adsabs.harvard.edu/abs/2010MNRAS.404..575L} {404, 575}

\bibitem[\protect\citeauthoryear{{Lotz}, {Jonsson}, {Cox}, {Croton}, {Primack},
  {Somerville}  \& {Stewart}}{{Lotz} et~al.}{2011}]{Lotz2011}
{Lotz} J.~M.,  {Jonsson} P.,  {Cox} T.~J.,  {Croton} D.,  {Primack} J.~R.,
  {Somerville} R.~S.,   {Stewart} K.,  2011, \mn@doi [\apj]
  {10.1088/0004-637X/742/2/103}, \href
  {http://adsabs.harvard.edu/abs/2011ApJ...742..103L} {742, 103}

\bibitem[\protect\citeauthoryear{{Man}, {Zirm}  \& {Toft}}{{Man}
  et~al.}{2016}]{Man2016}
{Man} A. W.~S.,  {Zirm} A.~W.,   {Toft} S.,  2016, \mn@doi [\apj]
  {10.3847/0004-637X/830/2/89}, \href
  {https://ui.adsabs.harvard.edu/abs/2016ApJ...830...89M} {830, 89}

\bibitem[\protect\citeauthoryear{{Mantha} et~al.,}{{Mantha}
  et~al.}{2018}]{mantha2018}
{Mantha} K.~B.,  et~al., 2018, \mn@doi [\mnras] {10.1093/mnras/stx3260}, \href
  {http://adsabs.harvard.edu/abs/2018MNRAS.475.1549M} {475, 1549}

\bibitem[\protect\citeauthoryear{{Mantha} et~al.,}{{Mantha}
  et~al.}{2019}]{Mantha2019}
{Mantha} K.~B.,  et~al., 2019, \mn@doi [\mnras] {10.1093/mnras/stz872}, \href
  {https://ui.adsabs.harvard.edu/abs/2019MNRAS.486.2643M} {486, 2643}

\bibitem[\protect\citeauthoryear{{Marian} et~al.,}{{Marian}
  et~al.}{2020}]{Marian2020}
{Marian} V.,  et~al., 2020, \mn@doi [\apj] {10.3847/1538-4357/abbd3e}, \href
  {https://ui.adsabs.harvard.edu/abs/2020ApJ...904...79M} {904, 79}

\bibitem[\protect\citeauthoryear{{Martin}, {Kaviraj}, {Devriendt}, {Dubois}  \&
  {Pichon}}{{Martin} et~al.}{2018}]{Martin2018}
{Martin} G.,  {Kaviraj} S.,  {Devriendt} J.~E.~G.,  {Dubois} Y.,   {Pichon} C.,
   2018, \mn@doi [\mnras] {10.1093/mnras/sty1936}, \href
  {https://ui.adsabs.harvard.edu/abs/2018MNRAS.480.2266M} {480, 2266}

\bibitem[\protect\citeauthoryear{{Moody}, {Romanowsky}, {Cox}, {Novak}  \&
  {Primack}}{{Moody} et~al.}{2014}]{Moody2014}
{Moody} C.~E.,  {Romanowsky} A.~J.,  {Cox} T.~J.,  {Novak} G.~S.,   {Primack}
  J.~R.,  2014, \mn@doi [\mnras] {10.1093/mnras/stu1444}, \href
  {https://ui.adsabs.harvard.edu/abs/2014MNRAS.444.1475M} {444, 1475}

\bibitem[\protect\citeauthoryear{{Moster}, {Somerville}, {Maulbetsch}, {van den
  Bosch}, {Macci{\`o}}, {Naab}  \& {Oser}}{{Moster} et~al.}{2010}]{Moster2010}
{Moster} B.~P.,  {Somerville} R.~S.,  {Maulbetsch} C.,  {van den Bosch} F.~C.,
  {Macci{\`o}} A.~V.,  {Naab} T.,   {Oser} L.,  2010, \mn@doi [\apj]
  {10.1088/0004-637X/710/2/903}, \href
  {https://ui.adsabs.harvard.edu/abs/2010ApJ...710..903M} {710, 903}

\bibitem[\protect\citeauthoryear{{Moster}, {Naab}  \& {White}}{{Moster}
  et~al.}{2013}]{Moster2013}
{Moster} B.~P.,  {Naab} T.,   {White} S. D.~M.,  2013, \mn@doi [\mnras]
  {10.1093/mnras/sts261}, \href
  {https://ui.adsabs.harvard.edu/abs/2013MNRAS.428.3121M} {428, 3121}

\bibitem[\protect\citeauthoryear{{Moster}, {Naab}  \& {White}}{{Moster}
  et~al.}{2018}]{Moster2018}
{Moster} B.~P.,  {Naab} T.,   {White} S.~D.~M.,  2018, \mn@doi [\mnras]
  {10.1093/mnras/sty655}, \href
  {http://adsabs.harvard.edu/abs/2018MNRAS.477.1822M} {477, 1822}

\bibitem[\protect\citeauthoryear{{Moster}, {Naab}  \& {White}}{{Moster}
  et~al.}{2020}]{Moster2020}
{Moster} B.~P.,  {Naab} T.,   {White} S. D.~M.,  2020, \mn@doi [\mnras]
  {10.1093/mnras/staa3019}, \href
  {https://ui.adsabs.harvard.edu/abs/2020MNRAS.499.4748M} {499, 4748}

\bibitem[\protect\citeauthoryear{{Mundy}, {Conselice}, {Duncan}, {Almaini},
  {H{\"a}u{\ss}ler}  \& {Hartley}}{{Mundy} et~al.}{2017}]{Mundy2017}
{Mundy} C.~J.,  {Conselice} C.~J.,  {Duncan} K.~J.,  {Almaini} O.,
  {H{\"a}u{\ss}ler} B.,   {Hartley} W.~G.,  2017, \mn@doi [\mnras]
  {10.1093/mnras/stx1238}, \href
  {https://ui.adsabs.harvard.edu/abs/2017MNRAS.470.3507M} {470, 3507}

\bibitem[\protect\citeauthoryear{{Naab} et~al.,}{{Naab}
  et~al.}{2014}]{Naab2014}
{Naab} T.,  et~al., 2014, \mn@doi [\mnras] {10.1093/mnras/stt1919}, \href
  {https://ui.adsabs.harvard.edu/abs/2014MNRAS.444.3357N} {444, 3357}

\bibitem[\protect\citeauthoryear{{Nevin}, {Blecha}, {Comerford}  \&
  {Greene}}{{Nevin} et~al.}{2019}]{Nevin2019}
{Nevin} R.,  {Blecha} L.,  {Comerford} J.,   {Greene} J.,  2019, \mn@doi [\apj]
  {10.3847/1538-4357/aafd34}, \href
  {https://ui.adsabs.harvard.edu/abs/2019ApJ...872...76N} {872, 76}

\bibitem[\protect\citeauthoryear{{O'Leary}, {Moster}, {Naab}  \&
  {Somerville}}{{O'Leary} et~al.}{2021}]{Oleary2021}
{O'Leary} J.~A.,  {Moster} B.~P.,  {Naab} T.,   {Somerville} R.~S.,  2021,
  \mn@doi [\mnras] {10.1093/mnras/staa3746}, \href
  {https://ui.adsabs.harvard.edu/abs/2021MNRAS.501.3215O} {501, 3215}

\bibitem[\protect\citeauthoryear{{Pfister}, {Dotti}, {Laigle}, {Dubois}  \&
  {Volonteri}}{{Pfister} et~al.}{2020}]{Pfister2020}
{Pfister} H.,  {Dotti} M.,  {Laigle} C.,  {Dubois} Y.,   {Volonteri} M.,  2020,
  \mn@doi [\mnras] {10.1093/mnras/staa227}, \href
  {https://ui.adsabs.harvard.edu/abs/2020MNRAS.493..922P} {493, 922}

\bibitem[\protect\citeauthoryear{{Pillepich} et~al.,}{{Pillepich}
  et~al.}{2018}]{Pillepich2018}
{Pillepich} A.,  et~al., 2018, \mn@doi [\mnras] {10.1093/mnras/stx2656}, \href
  {https://ui.adsabs.harvard.edu/abs/2018MNRAS.473.4077P} {473, 4077}

\bibitem[\protect\citeauthoryear{{Planck Collaboration}}{{Planck
  Collaboration}}{2016}]{Planck2016}
{Planck Collaboration} 2016, \mn@doi [\aap] {10.1051/0004-6361/201525830},
  \href {http://adsabs.harvard.edu/abs/2016A%26A...594A..13P} {594, A13}

\bibitem[\protect\citeauthoryear{{Ragan-Kelley}, {Perez}, {Granger}, {Kluyver},
  {Ivanov}, {Frederic}  \& {Bussonnier}}{{Ragan-Kelley}
  et~al.}{2014}]{Ragan-Kelley:2014aa}
{Ragan-Kelley} M.,  {Perez} F.,  {Granger} B.,  {Kluyver} T.,  {Ivanov} P.,
  {Frederic} J.,   {Bussonnier} M.,  2014, in AGU Fall Meeting Abstracts. pp
  H44D--07

\bibitem[\protect\citeauthoryear{{Scarlata} et~al.,}{{Scarlata}
  et~al.}{2007}]{Scarlata2007}
{Scarlata} C.,  et~al., 2007, \mn@doi [\apjs] {10.1086/516582}, \href
  {https://ui.adsabs.harvard.edu/abs/2007ApJS..172..406S} {172, 406}

\bibitem[\protect\citeauthoryear{{Schaye} et~al.,}{{Schaye}
  et~al.}{2015}]{Schaye2015}
{Schaye} J.,  et~al., 2015, \mn@doi [\mnras] {10.1093/mnras/stu2058}, \href
  {https://ui.adsabs.harvard.edu/abs/2015MNRAS.446..521S} {446, 521}

\bibitem[\protect\citeauthoryear{{Sharma} et~al.,}{{Sharma}
  et~al.}{2021}]{Sharma2021}
{Sharma} R.~S.,  et~al., 2021, arXiv e-prints, \href
  {https://ui.adsabs.harvard.edu/abs/2021arXiv210101729S} {p. arXiv:2101.01729}

\bibitem[\protect\citeauthoryear{{Shi}, {Rieke}, {Lotz}  \&
  {Perez-Gonzalez}}{{Shi} et~al.}{2009}]{Shi2009}
{Shi} Y.,  {Rieke} G.,  {Lotz} J.,   {Perez-Gonzalez} P.~G.,  2009, \mn@doi
  [\apj] {10.1088/0004-637X/697/2/1764}, \href
  {https://ui.adsabs.harvard.edu/abs/2009ApJ...697.1764S} {697, 1764}

\bibitem[\protect\citeauthoryear{{Snyder}, {Lotz}, {Rodriguez-Gomez},
  {Guimar{\~a}es}, {Torrey}  \& {Hernquist}}{{Snyder}
  et~al.}{2017}]{Snyder2017}
{Snyder} G.~F.,  {Lotz} J.~M.,  {Rodriguez-Gomez} V.,  {Guimar{\~a}es} R.
  d.~S.,  {Torrey} P.,   {Hernquist} L.,  2017, \mn@doi [\mnras]
  {10.1093/mnras/stx487}, \href
  {https://ui.adsabs.harvard.edu/abs/2017MNRAS.468..207S} {468, 207}

\bibitem[\protect\citeauthoryear{{Somerville}, {Hopkins}, {Cox}, {Robertson}
  \& {Hernquist}}{{Somerville} et~al.}{2008}]{Somerville2008}
{Somerville} R.~S.,  {Hopkins} P.~F.,  {Cox} T.~J.,  {Robertson} B.~E.,
  {Hernquist} L.,  2008, \mn@doi [\mnras] {10.1111/j.1365-2966.2008.13805.x},
  \href {https://ui.adsabs.harvard.edu/abs/2008MNRAS.391..481S} {391, 481}

\bibitem[\protect\citeauthoryear{{Springel}}{{Springel}}{2005}]{gadget2}
{Springel} V.,  2005, \mn@doi [\mnras] {10.1111/j.1365-2966.2005.09655.x},
  \href {http://adsabs.harvard.edu/abs/2005MNRAS.364.1105S} {364, 1105}

\bibitem[\protect\citeauthoryear{{Steinborn}, {Hirschmann}, {Dolag}, {Shankar},
  {Juneau}, {Krumpe}, {Remus}  \& {Teklu}}{{Steinborn}
  et~al.}{2018}]{Steinborn2018}
{Steinborn} L.~K.,  {Hirschmann} M.,  {Dolag} K.,  {Shankar} F.,  {Juneau} S.,
  {Krumpe} M.,  {Remus} R.-S.,   {Teklu} A.~F.,  2018, \mn@doi [\mnras]
  {10.1093/mnras/sty2288}, \href
  {https://ui.adsabs.harvard.edu/abs/2018MNRAS.481..341S} {481, 341}

\bibitem[\protect\citeauthoryear{{Ventou} et~al.,}{{Ventou}
  et~al.}{2017}]{Ventou2017}
{Ventou} E.,  et~al., 2017, \mn@doi [\aap] {10.1051/0004-6361/201731586}, \href
  {https://ui.adsabs.harvard.edu/abs/2017A&A...608A...9V} {608, A9}

\bibitem[\protect\citeauthoryear{{Ventou} et~al.,}{{Ventou}
  et~al.}{2019}]{Ventou2019}
{Ventou} E.,  et~al., 2019, \mn@doi [\aap] {10.1051/0004-6361/201935597}, \href
  {https://ui.adsabs.harvard.edu/abs/2019A&A...631A..87V} {631, A87}

\bibitem[\protect\citeauthoryear{{Virtanen} et~al.,}{{Virtanen}
  et~al.}{2020}]{Virtanen:2020aa}
{Virtanen} P.,  et~al., 2020, \mn@doi [Nature Methods]
  {https://doi.org/10.1038/s41592-019-0686-2}, \href {https://rdcu.be/b08Wh}
  {17, 261}

\bibitem[\protect\citeauthoryear{{Vogelsberger} et~al.,}{{Vogelsberger}
  et~al.}{2014}]{Vogelsberger2014b}
{Vogelsberger} M.,  et~al., 2014, \mn@doi [\mnras] {10.1093/mnras/stu1536},
  \href {https://ui.adsabs.harvard.edu/abs/2014MNRAS.444.1518V} {444, 1518}

\bibitem[\protect\citeauthoryear{{Wen} \& {Zheng}}{{Wen} \&
  {Zheng}}{2016}]{Wen2016}
{Wen} Z.~Z.,  {Zheng} X.~Z.,  2016, \mn@doi [\apj]
  {10.3847/0004-637X/832/1/90}, \href
  {https://ui.adsabs.harvard.edu/abs/2016ApJ...832...90W} {832, 90}

\bibitem[\protect\citeauthoryear{{Yoon} \& {Lim}}{{Yoon} \&
  {Lim}}{2020}]{Yoon2020}
{Yoon} Y.,  {Lim} G.,  2020, \mn@doi [\apj] {10.3847/1538-4357/abc621}, \href
  {https://ui.adsabs.harvard.edu/abs/2020ApJ...905..154Y} {905, 154}

\bibitem[\protect\citeauthoryear{{van der Walt}, {Colbert}  \&
  {Varoquaux}}{{van der Walt} et~al.}{2011}]{van-der-Walt:2011aa}
{van der Walt} S.,  {Colbert} S.~C.,   {Varoquaux} G.,  2011, \mn@doi [CiSE]
  {10.1109/MCSE.2011.37}, \href
  {https://ui.adsabs.harvard.edu/abs/2011CSE....13b..22V} {13, 22}

\makeatother
\end{thebibliography}



\appendix





\bsp	
\label{lastpage}
\end{document}